# Giant viscoelasticity near Mott criticality in $PbCrO_3$ with large lattice anomalies

Shanmin Wang*, Jian Chen, Liusuo Wu, and Yusheng Zhao

*Department of Physics & Academy for Advanced Interdisciplinary Studies, Southern University of Science & Technology, Shenzhen, Guangdong, 518055, China*

**Email:* wangsm@sustech.edu.cn

## Abstract

Coupling of charge and lattice degrees of freedom in materials can produce intriguing electronic phenomena, such as conventional superconductivity where the electrons are mediated by lattice for creating supercurrent. The Mott transition, which is a source for many fascinating emergent behaviors, is originally thought to be driven solely by correlated electrons with an Ising criticality. Recent studies on the known Mott systems have shown that the lattice degree of freedom is also at play, giving rise to either Landau or unconventional criticality. However, the underlying coupling mechanism of charge and lattice degrees of freedom around the Mott critical endpoint remains elusive, leading to difficulties in understanding the associated Mott physics. Here we report a study of Mott transition in cubic $PbCrO_3$ by measuring the lattice parameter, using high-pressure x-ray diffraction techniques. The Mott criticality in this material is revealed with large lattice anomalies, which is governed by giant viscoelasticity that presumably results from a combination of lattice elasticity and electron viscosity. Because of the viscoelastic effect, the lattice of this material behaves peculiarly near the critical endpoint, inconsistent with any existing university classes. We argue that the viscoelasticity may play as a hidden degree of freedom behind the Mott criticality.

**Keywords:** *Mott transition, $PbCrO_3$, viscoelasticity, Mott criticality, correlated system*

## Introduction

The correlated electrons in materials can condense into new forms of collective states by varying pressure (P) and temperature (T), and they are a key to many emergent phenomena, such as unconventional superconductivity[1], spin-liquid phases[2], and quantum criticality[1-3]. Understanding of these phenomena is closely linked to the Mott insulator-metal transition that is mainly characterized by a first-order isostructural transition ended at a second-order critical endpoint in the associated P-T space [4]. Critical properties of a Mott system are expected to be described in the framework of the scaling theory of a known universality class[5,6], allowing gaining insights into collective nature of the system.

The first experimental determination of Mott critical behaviors comes to the Cr-doped $V_2O_3$ by conductivity measurements, revealing an Ising criticality[6], and it can be understood in terms of the classical liquid-gas transition, indicating a purely electronic system[7,8]. However, slightly away from the endpoint of this system, the Landau criticality is favored, due to the effect of lattice deformation on its electronic system [7]. Intriguingly, an unconventional criticality in an organic salt $\kappa$-(BEDT-TTF)$_2$Cu[N(CN)$_2$]Cl (denoted by $\kappa$-Cl hereafter) was also found[9,10], which is distinct from any existing universality classes. According to Zacharias *et al.*[7], incorporating the lattice elasticity into a purely electronic critical state can lead to the suppression of microscopic charge fluctuation by the long-ranged shear stress, which explains the crossover from Ising to either Landau or unconventional criticality with a more rapid ordering of the charge degree of freedom[9]. However, the underlying charge-lattice coupling mechanism for the Mott criticality is still unclear.

Intuitively, the lattice seemingly plays as a glue for mediating electrons to form new collective states around the critical endpoint, similar to the mechanism for forming Cooper pairs in conventional superconductors[11]. Such new electronic states should have viscosity due to the frequent electron-electron collisions, as the thermally-induced scattering by the lattice is completely suppressed. In fact, the electron viscosity has been revealed by a surge of recent experiments[12-15], although it was originally predicted by Landau[12]. The thus-produced electron viscosity combined with lattice elasticity would produce viscoelasticity in the associated materials. Materials with viscoelastic properties are expected to exhibit many mechanical anomalies[16-18], which can explain the Mott critical behaviors.

The lattice strain can thus provide a straightforward probe for exploring the viscoelastic effect. Using both the ultrasound and dilatometric measurements, a large elastic softening in Cr-doped $V_2O_3$[19,20] and anomalous thermal expansions in $\kappa$-Cl were previously reported around their critical endpoints[21]. A recent isothermal dilatometric experiment on κ-Cl has led to a successful study of its Mott criticality by monitoring the macroscopic sample length[8], but it suffers from complicated data analysis as multiple strain components are involved. By contrast, high-P x-ray diffraction (XRD) combined with a precise temperature control is a powerful tool to look into the Mott transition by probing lattice parameters under varying P-T conditions[22,23]. The expected critical elastic anomalies can thus be explored by volumetric (V) analysis. However, the known Mott systems with low critical pressures (i.e., below 1 GPa) are unfavorable for high-P XRD experiment. In this regard, as a recently-identified Mott system $PbCrO_3$ is suitable with a moderate critical pressure of ~3 GPa[24]. Besides, conflicting mechanisms of charge disproportionation[25] and melting of charge glass [26] were also proposed for explaining pressure-induced isostructural transition in this oxide [27], and ambiguities of its magnetic, electronic, and ferroelectric properties still remain [24-34], calling for more experimental efforts. Here we present a systematic high-P study of $PbCrO_3$ with a focus on its lattice Mott criticality (see Supplemental Material for experimental details), leading to the discovery of giant viscoelasticity around its critical endpoint. Based on the viscoelastic effect, the observed elastic anomalies and unconventional lattice criticality can be well interpreted, shedding light on the underlying mechanism of Mott transition.

## Results and discussion

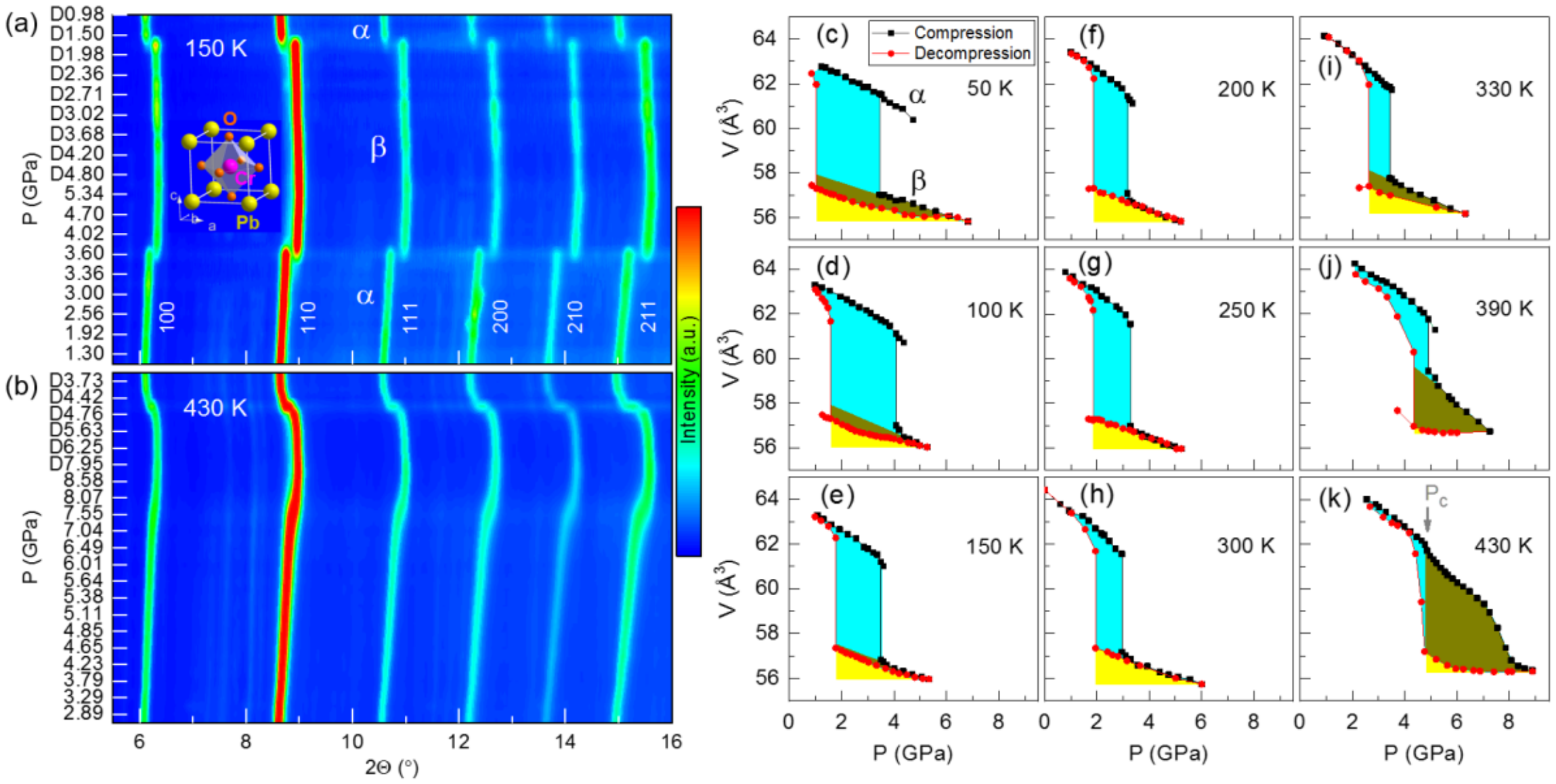

**Fig. 1.** Isothermal high-P XRD patterns and P-V data for $PbCrO_3$. (**a**)-(**b**) Contour plots of high-P XRD patterns collected upon both compression and decompression ("D") at 150 K and 430 K, respectively. Inset is the unit cell. The incident x-ray wavelength is $\lambda$ = 0.4246 Å. (**c**)-(**k**) Isothermal P-V data. In each panel, the cyan, dark yellow, and yellow shaded regions correspond to the phase-transition hysteresis loss, viscoelastic dissipation, and recoverable elastic energy, respectively. The error bars are too small to show.

High-P XRD patterns of $PbCrO_3$ taken during isothermal loading cycles are shown in Figs. 1(a)-1(b) (see Figs. S1-S11 for detailed data analysis and structural refinements[35]). For simplicity, the low-P insulating and high-P metallic phases are hereafter denoted with α and β, respectively[24]. Below 390 K, the pressure-induced α→β transition is primarily characterized by a remarkable shift of each the involved diffraction peak without the appearance of additional peaks, suggesting a first-order isostructural transition (Fig. 1(a)), consistent with previous reports[24,25,27]. However, at 430 K the forward α-β transition becomes a second order without sharp peak shifts (Fig. 1(b)), and the transition pressure is hardly discerned from the peak shift. Upon decompression, the reverse transition can still be identified by peak shifts at ~4.8 GPa, although the shifts are continuous. Besides, a prominent peak broadening is occurred above 4 GPa and reaches maximum around 7.6 GPa with a rapid lattice softening (Fig. 1(b) and Fig. S11), due to the viscoelastic effect as discussed later.

Figs. 1(c)-1(k) present the obtained high-P isothermal volume data of $PbCrO_3$. In each loading cycle, a hysteresis loop can be defined by both the forward and reverse thresholds. Clearly, the loop area is equal to an energy dissipation related to the structural transition kinetics[36,37]. With increasing temperatures, the pressure hysteresis progressively decreases, especially above 300 K, which eventually vanishes at a critical temperature ($T_c$) of ~430 K (Fig. 1(k)); similar hysteresis can also be identified from the variation of phase fraction (Fig. S8). In each isothermal P-V line of β-$PbCrO_3$ below 150 K and above 300 K, a deviation appears between the compression and decompression processes, leading to an extra P-V loop, a characteristic of viscoelastic behaviors[38]. Note that the viscoelastic effect can only appear in the β phase with delocalized electrons and likely results from a combination of electron viscosity and lattice elasticity. The thus-produced extra loop area corresponds to an additional energy dissipation, due to the viscoelastic deformation. Recoverable elastic energy in β-$PbCrO_3$ can be integrated from its decompression P-V

curve. With decreasing temperature below 150 K, such viscoelastic loop is enlarged with reducing recoverable elasticity, indicating enhanced viscoelasticity (Figs. 1(c)-1(d)). Interestingly, the loop area promptly grows above 300 K. At $T_c$, the viscoelastic deformation prevails, leading to a huge amount of energy dissipation and nearly zero recoverable energy. In contrast, at 200 - 300 K the β phase exhibits a purely elastic nature without involvement of viscoelastic loop (Figs. 1(f)-1(h)).

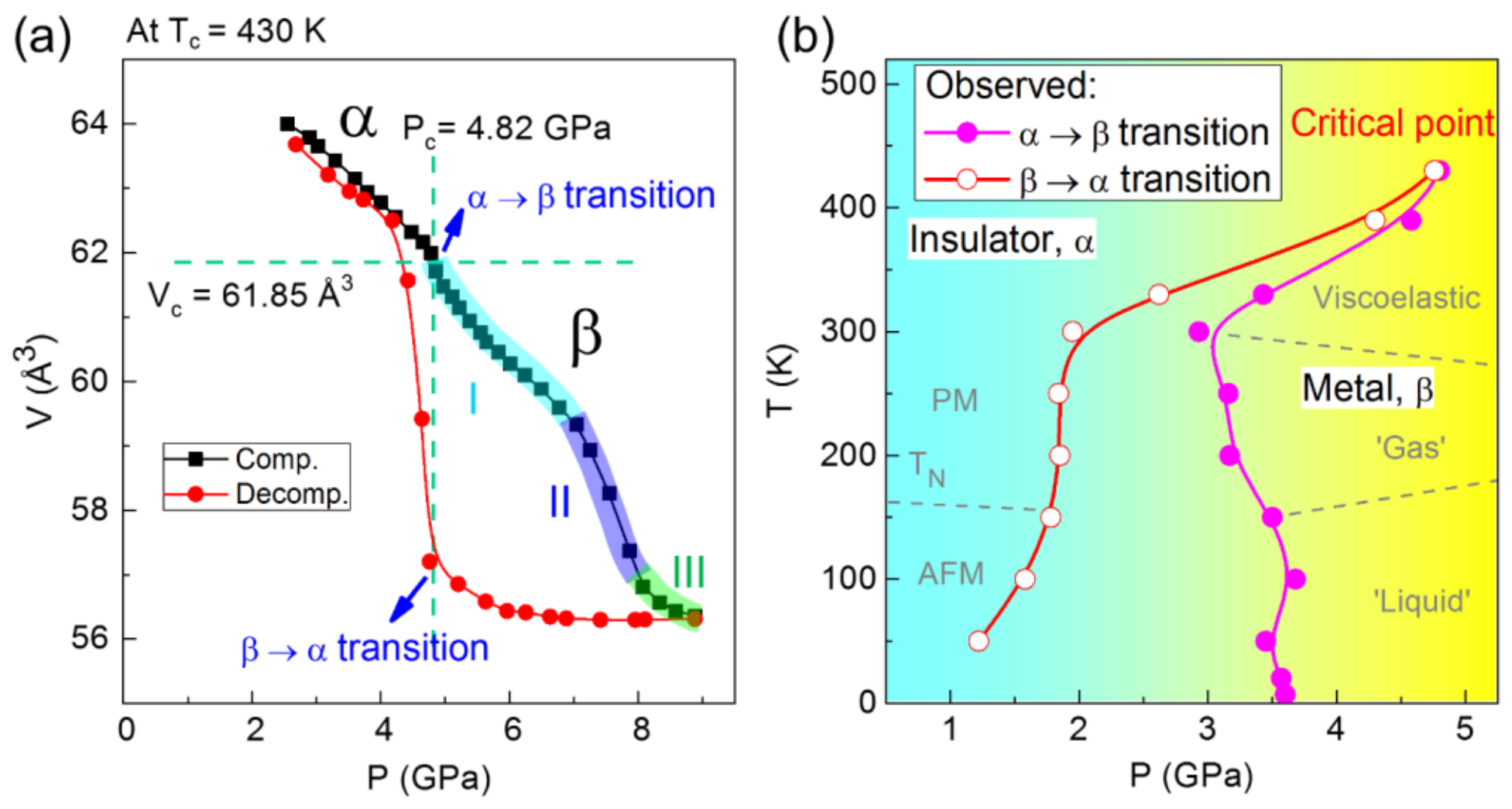

**Fig. 2.** Lattice Mott criticality near the endpoint ($P_c$, $T_c$) and phase diagram of $PbCrO_3$. (**a**) P-V data at $T_c$ = 430 K. Compression P-V line of β can be divided into three different regions of I, II, and III, as highlighted in cyan, blue, and green, respectively. (**b**) P-T phase diagram. The two spinodal lines are defined by the forward and reverse α-β thresholds. The delocalized electrons in β-$PbCrO_3$ have three different states of 'liquid', 'gas', and viscoelastic state. Paramagnetic (PM)-to-antiferromagnetic (AFM) transition line ($T_N$) determined by high-P neutron diffraction will be published elsewhere.

On a close look at the forward transition at $T_c$ (Fig. 2(a)), a remarkably different lattice behavior starts from 4.82 GPa. Coincidently, the reverse transition occurs at the nearly same pressure during decompression, indicating a critical pressure ($P_c$) of ~4.82 GPa. The thus-determined Mott critical endpoint for $PbCrO_3$ is at $T_c \approx 430$ K and $P_c \approx 4.82$ GPa. Because of the viscoelastic effect, the volume reduction in forward transition is suppressed, resulting in a second-order transition. However, a prompt volume contraction above 7 GPa may indicate a rapidly varying viscoelasticity with anomalous peak broadening, especially around 7.6 GPa (Fig. 1(b)). Above 8 GPa, the lattice of β-$PbCrO_3$ has a normal stiffness (i.e., $B_0 \approx 113$), close to that at 300 K with pure elasticity, whereas it remains nearly invariant during decompression till the reverse transition, implying viscoelasticity-induced anomalies in elasticity. On the removal of pressure, β-$PbCrO_3$ can fully recover to its original phase with characteristics of anelasticity, a special case of viscoelasticity[39].

Based on these experimental observations, a P-T phase diagram of $PbCrO_3$ is depicted in Fig. 2(b). Apparently, the P-T space is mainly divided into three regions by two spinodal lines, including insulating α, metallic β, and two-phase coexistence regions. The two spinodal lines are constrained by first-order transition lines, terminating in a critical endpoint ($T_c$, $P_c$), beyond which the transition is second order. The similar phase diagrams have been observed in other Mott systems (e.g., Cr-doped $V_2O_3$[4,6] and $\kappa$-Cl[40,41]). In the β region, the delocalized electrons may have three distinct states of 'liquid', 'gas', and viscoelastic state as discussed later; such a picture is reminiscent of water with liquid, gas, and supercritical fluid states under P-T conditions[5]. Interestingly, the shape of two spinodal lines is seemingly affected by these electronic states. In particular, the sharpening of two-phase coexistence region coincidently occurs in the vicinity of appearance of viscoelasticity above 300 K. Besides, the super-large viscoelasticity is expected at ($T_c$, $P_c$). Thus, coupling of lattice elasticity and electron viscosity in β-$PbCrO_3$ could generate giant viscoelasticity, which ultimately governs the Mott critical properties.

For a Mott system, the critical exponents of observable quantities around the critical endpoint is important, based on which the transition can be classified and understood within the framework of the scaling theory of classical systems[5]. To extract the Mott critical exponents for $PbCrO_3$, the relative lattice densities, $|\rho - \rho_c|$, of both the insulating and metallic phases at 430 K are plotted in Fig. 3(a) as a function of pressure, $|P - P_c|$, obeying the same power law $|P - P_c|^{1/\delta}$ with $\delta \approx 4/3$. On the border of coexistence region, the variation of relative density of the metallic phase ($\rho_{metal} - \rho_c$) against temperature ($T_c$ - T) is shown in Fig. 3(b). Using the scaling relation $(T_c - T)^{\beta}$, the best fit can be achieved for only the high-T region (i.e., 300 – 430 K) with $\beta \approx 3/5$. Therefore, the obtained critical exponents are $(\delta, \beta) \approx (4/3, 3/5)$ (Fig. 3(c)), inconsistent with those of existing universality classes (i.e., $\delta \geq 3$ and $\beta \leq 1/2$) bounded by the mean-field values of $(\delta, \beta) = (3, 1/2)$ or the well-studied Mott systems (e.g., $(V_{1-x}Cr_x)_2O_3$[6] and organic $\kappa$-Cl[9,10]).

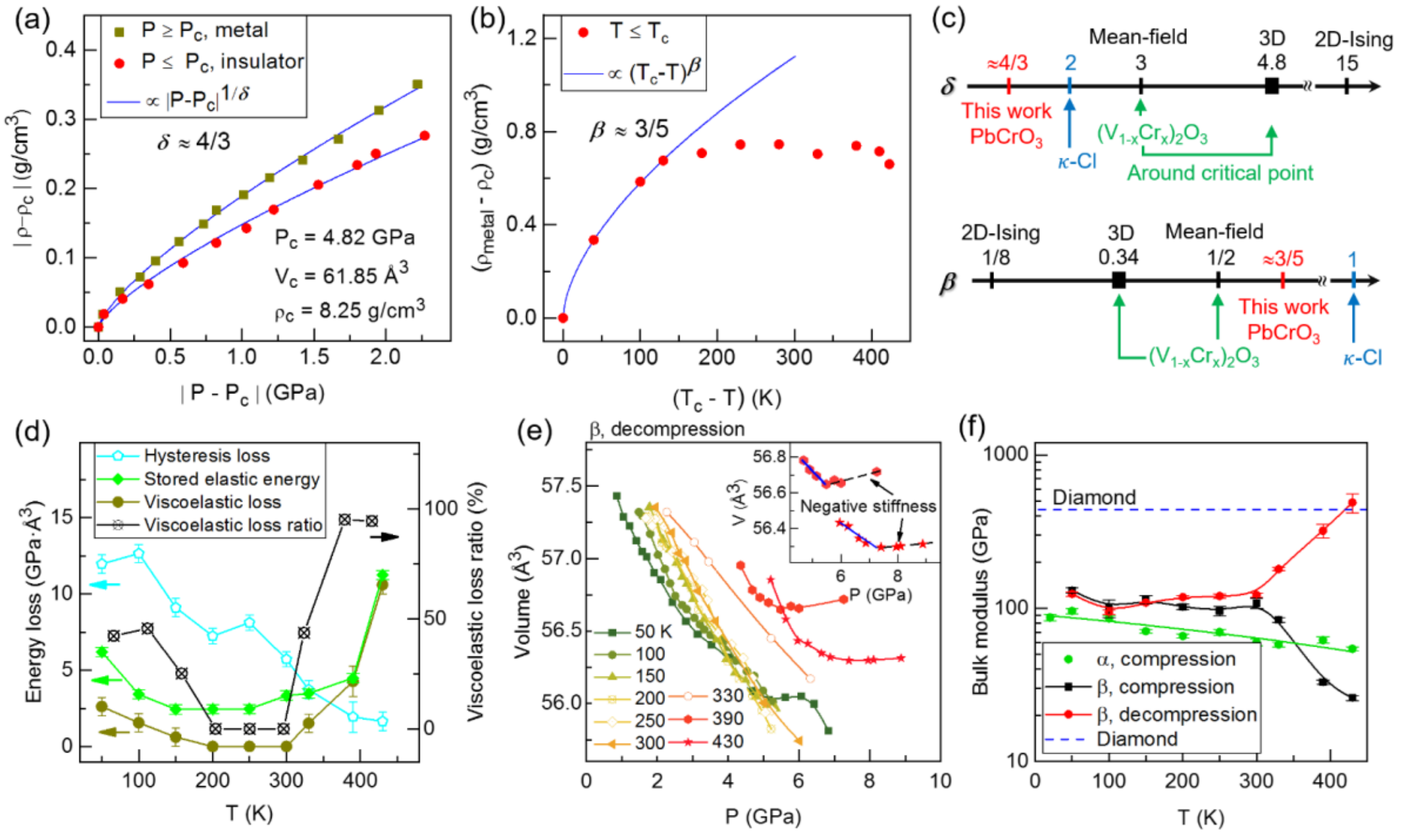

**Fig. 3.** Mott critical behaviors, energy dissipation, and viscoelastic stiffness. (**a**) Lattice density vs. pressure. (**b**) Lattice density vs. temperature for metallic phase on the border of coexistence region. (**c**) Comparison of critical exponents of $PbCrO_3$ with those of universality (e.g., mean-field, 2D-Ising, and 3D models[9]). Critical exponents of $(V_{1-x}Cr_x)_2O_3$ and $\kappa$-Cl are added[6,9]. (**d**) Total elastic energy storage in $\beta$-$PbCrO_3$, viscoelastic and hysteresis dissipations, and the ratio of viscoelastic loss to the stored elastic energy *vs.* temperature derived from integrations of the P-V data (Figs. 1(c)-1(k) and Fig. S12). The total elastic energy is the sum of viscoelastic and recoverable elastic energies. (**e**) Isothermal P-V data for $\beta$-$PbCrO_3$ collected during decompression. Inset is an enlarged portion of the cases at 390 K and 430 K. Each blue line represents

a fit of the associated isothermal P-V data to a 2nd-order Birch-Murnaghan EoS. (**f**) Isothermal bulk moduli for both α- and β-$PbCrO_3$.

Remarkably, $PbCrO_3$ exhibits unconventional critical behaviors, which can be understood in terms of the viscoelastic effect, indicating that the real order parameter of this system may be viscoelasticity, rather than the pure charge or lattice degree of freedom. The viscoelasticity is closely related to the high-P metallic phase and should originate from a synthesis of electron viscosity and lattice elasticity. The relative ratio between these two components could be changed in different systems, leading to distinctly different critical behaviors. From this viewpoint, the viscous component in $(V_{1-x}Cr_x)_2O_3$ is likely dominated, because its critical behaviors obey the mean-field and 3D models (Fig. 3(c))[6]. However, the Mott criticalities in $PbCrO_3$ and $\kappa$-Cl probably involve large fractions of elastic component[9,10], which results in unconventional critical exponents far beyond the limit of conventional classes, giving rise to more rapid orderings of order parameters.

To quantitatively explore the viscoelastic effect, we derive the associated energy dissipations and the totally-stored and recoverable elastic energies in the β phase by integrating the P-V areas (Fig. 3(d) and Fig. S12). As expected, the hysteresis-induced energy dissipation progressively reduces to zero, as the $T_c$ is approached. By contrast, the viscoelastic dissipation is small at 50 K and gradually decreases to zero at 200 - 300 K, followed by a rapid increase at 330 - 430 K. In fact, the viscoelasticity can be quantified by the ratio of viscoelastic energy to total elastic energy storage in β-$PbCrO_3$ during compression. Around $T_c$, nearly all the stored elastic energy is dissipated by the viscoelastic deformation, producing a giant viscoelastic effect. During decompression, the lattice of β-$PbCrO_3$ exhibit complex responses to these energy variations (Fig. 3(e) and Figs. S13-S14). At 50 K, clear lattice anomalies occur at 3 - 7 GPa and progressively diminish at 200 K. In the 200 - 300 K range, the isothermal volume data has a similar linear variation with pressure, which can be well described by the Birch-Murnaghan equation of state (EoS)[42], in striking contrast to that around $T_c$ with large lattice anomalies.

Considering no structural change is involved, these phenomena can only be associated with delocalized electrons in β-$PbCrO_3$, which could have distinct collective states with varying temperature, hence different viscosities[12]. It is speculated that a liquid-like state occurs below 150 K with certain viscosity, because the electrons are weakly scattered by the lattice at low temperatures. At 200 - 300 K, the electrons may be frequently scattered by the lattice, so that the electron-electron collisions are hindered, leading to the formation of a non-viscous gas-like state. In this case, the interplay between the electrons and lattice elasticity is weak, which is responsible for its normal elastic behaviors (Figs. 1(f) - 1(h)). However, the viscoelastic effect is counter-intuitively increased above 300 K, inferring an unusual mechanism for the formation of large electron viscosity. As mentioned earlier, the delocalized electrons are likely glued by the lattice to form viscous flow, and the viscosity increases upon approaching to $T_c$, which may even lead to superviscosity at the critical endpoint. Such viscous electrons can react with the lattice to form viscoelasticity in β-$PbCrO_3$ in the 50 - 150 K and 330 - 430 K ranges, causing lattice anomalies. Above 390 K, because of the super-large viscoelasticity, the negative stiffness occurs in the β phase with positive slopes at the early stage of decompression (i.e., volume shrinking upon decompression) (Fig. 3(e)).

In principle, a material with negative stiffness is unstable, unless it is pre-deformed (i.e., containing stored energy) and constrained by a soft surrounding matrix to form inclusion-matrix composite[16-18,43-45]. Negative stiffness can produce extreme physical properties in the composite including giant damping, time-dependent strain, and extremely high dynamic elastic modulus[18,46], as observed in the Sn-included $BaTiO_3$ and $VO_2$ composites[16,18]. Obviously, β-$PbCrO_3$ is a high-P metallic phase with a pre-strained volume; around the critical endpoint, the delocalized electrons would form a superviscous fluid that plays as a soft 'matrix', satisfying the prerequisites for producing negative stiffness. Besides, the expected extremely large dynamic bulk modulus ($B_0$) of $PbCrO_3$ are also observed (Fig. 3(f)). Below 300 K, the bulk modulus of β phase is levelled-off at ~120 GPa, irrespective of compression or decompression; however, above 300 K it increases exponentially and reaches ~490 GPa at $T_c$ during decompression (Fig. 3(f) and Fig. S14), which is even substantially greater than that of diamond (i.e., 440 GPa). For the cases obtained during compression, the trend is reversed and the bulk modulus is exponentially decreased (e.g., only ~26 GPa at $T_c$). These elastic anomalies can be well explained according to the viscoelastic effect. However, for the insulating α phase the derived $B_0$ decreases linearly without involving any anomalies (Fig. 3(f)), because there is no itinerant electron in this phase for producing electron viscosity. Due to the viscoelastic damping, the lattice strain in β-$PbCrO_3$ is dynamic in nature, which should be out of phase from the stress or applied pressure (e.g., the strain lags behind the stress), especially near the criticality. One of the remarkable consequences is that the sharp volume change is suppressed across the phase transition, resulting in a crossover from the first to second order transition. Last, using our recently-developed high-P technique, we performed dynamic high-P transport experiments[47], leading to the observation of a time-dependent effect of lattice contraction on the electrical resistance of β-$PbCrO_3$, which is probably associated with the viscoelasticity (Fig. S16).

In summary, by the high-P isothermal volumetric measurement, the Mott transition and lattice Mott criticality in $PbCrO_3$ are systematically investigated. The P-T phase diagram of this Mott system is well mapped out by two first-order-transition spinodal lines, terminating in a second-order critical endpoint at $T_c \approx 430$ K and $P_c \approx 4.82$ GPa. Close to the critical temperature, giant viscoelasticity is revealed in the high-P metallic phase, which is presumably a combined result of electrons viscosity and lattice elasticity. Based on the viscoelastic effect, a number of anomalous critical properties of this system can be well interpreted, including unusual Mott criticality, anomalous lattice behaviors, negative stiffness, and extremely high dynamic bulk modulus. In addition, our results would also be applicable to understanding the relevant problems, such as high-T superconductivity where the viscoelastic effect may be decisive.

## Acknowledgments

This work is supported by the National Natural Science Foundation of China (Grant No. 12174175), the Key Research Platforms and Research Projects of Universities in Guangdong Province (Grant No. 2018KZDXM062), the Guangdong Innovative & Entrepreneurial Research Team Program (No. 2016ZT06C279), the Shenzhen Peacock Plan (No. KQTD2016053019134356), the Shenzhen Development and Reform Commission Foundation for Shenzhen Engineering Research Center for Frontier Materials Synthesis at High Pressure, and the Research Platform for Crystal Growth & Thin-Film Preparation at SUSTech. High-P synchrotron XRD experiments were performed at the 16-BMD beamline of HPCAT/APS. We thank B. Hou, C. Park, and C. Kenney-Benson for their help on the set-up of high-P synchrotron experiments at HPCAT.

## Data availability

The data that support the findings of this study are available from the corresponding authors upon reasonable request.

## Competing interests

The authors declare no competing interests.

## Author information

*E-mail: wangsm@sustech.edu.cn (S. Wang)

## Supplemental Material

### Giant viscoelasticity near Mott criticality in $PbCrO_3$ with large lattice anomalies

Shanmin Wang*, Jian Chen, Liusuo Wu, and Yusheng Zhao

*Department of Physics & Academy for Advanced Interdisciplinary Studies, Southern University of Science & Technology, Shenzhen, Guangdong, 518055, China*

**Email: wangsm@sustech.edu.cn*

### Experimental details

$PbCrO_3$ sample was synthesized at 8 GPa and 1150 ºC for 30 min, using a solid-state reaction between PbO and $CrO_2$ [1]. Our high-P synchrotron XRD experiments were performed in diamond-anvil cells (DAC) at the 16BM-D beamline of APS with an incident x-ray wavelength of $\lambda$ = 0.4246 Å and a radiation energy of 29.2 keV, and the Mar3450 CCD detector was employed to collect the data. In each cycle of experiment, the sample powders were loaded in a sample hole (~150 um in diameter) of a pre-indented steel gasket (~40 in thickness). Low-T (i.e., 20 - 300 K) and high-T experiments (i.e., 330 - 430 K) were carried out in a liquid-flow helium cryostat and a high-T furnace, respectively, using the neon and silicon oil as pressure-transmitting media. The cell pressure was remotely controlled by a membrane system with an online ruby [2,3], showing a small pressure uncertainty within 0.03 GPa as checked before and after each the collection of data at desired P-T conditions. At each target temperature, the XRD data were collected during both the compression and decompression with a pressure step of ~0.2 GPa. The diffracted x-ray signals were collected for 5 min to obtain high-quality XRD data. The average time interval between successive two collections was ~5 min for changing and calibrating pressure and for sample positioning. The collected diffraction data were analyzed by integrating 2D images as a function of 2θ using the Dioptas program to obtain conventional, one-dimensional diffraction profiles. The thus-integrated one-dimensional high-P XRD data was analyzed using both the GSAS [4] and JADE (Materials Data, Inc.) programs to determine the lattice parameters. More experimental details can be found elsewhere [1].

To study the time-dependent variation of its electrical resistance against pressure for exploring the viscoelastic effect, we conducted high-P transport measurements for $PbCrO_3$, using our newly-established piezo-driven DAC techniques and a four-probe method [5]. A major advantage of this device is its specially-designed pressure control, allowing continuously varying pressure. Meanwhile, the pressure loading rate can be largely tunable from 0.5 to 120 GPa/min, which is important for dynamic experiments where a fast variation of pressure is required. By means of this device, the dynamics of collective behavior of charges can be explored under varying pressure at different time scales, based on the transport measurement. Provided that the increased viscosity of charges would result in an enhancement of viscoelasticity of a material, the deformation behavior of its lattice could be profoundly affected. As a result, the transport properties could be altered. Thus, in our experiment the sample was first compressed to different pressures of 1.2, 4.5, and 6.3 GPa, respectively. At each pressure, the cell was then heated to two different temperatures of 300 and 360 K. In order to study the time-dependent effect, the cell pressure was periodically oscillated with a small pressure amplitude of $\Delta P = 0.5$ GPa and various oscillation periods of 1, 2, 5, 10, 30, and 60 s. The similar experiment was also performed on a Mn-Cu alloy sample, which was used as a standard for the calibrations of cell pressure and experimental uncertainty. Note that the negligible viscoelasticity is expected in this alloy. Detailed electrodes preparation and pressure calibration can be found in ref. [5].

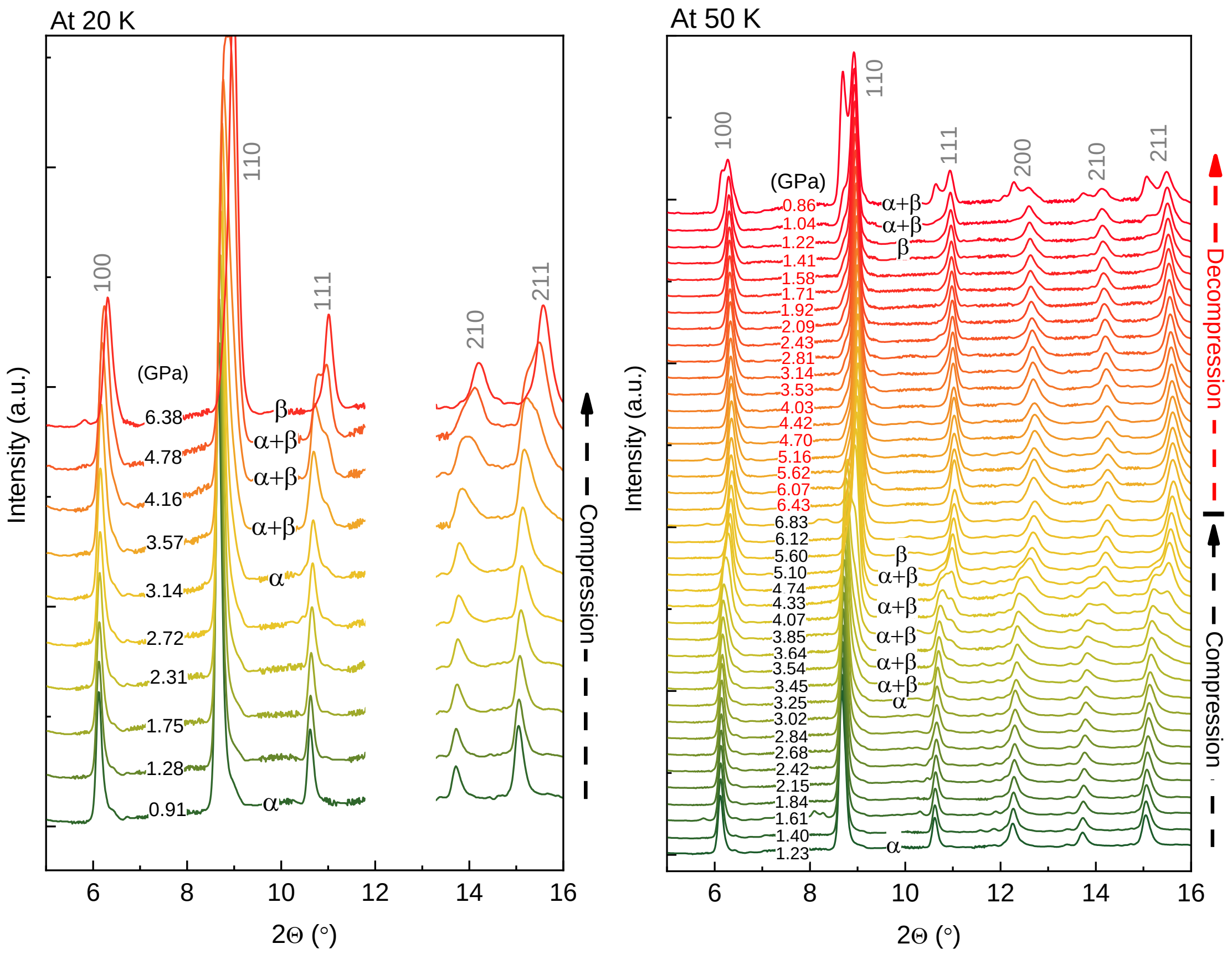

**Fig. S1.** High-P XRD patterns collected at 20 K (left panel) and 50 K (right panel). At 20 K, only the compression data were collected, because of the deteriorated hydrostatic pressure conditions, which leads to difficulties in the pressure control for taking decompression data. The broadened peak profiles at 20 K also suggest a non-hydrostatic pressure condition, because the neon pressure-transmitting medium likely becomes a solid below 20 K and at pressures above 2 GPa, which cannot maintain quasi-hydrostatic pressure conditions in the cell. Also noted is that the low-T experiments (e.g., 20 - 300 K) were conducted with neon as a pressure-transmitting medium. The incident synchrotron x-rays wavelength is $\lambda = 0.4246$ Å. The masked portion of patterns at 20 K corresponds to the strong diffraction peaks of the gasket system in the 12 - 13° range, because of a shift of beam center. The similar situations also occurred at 100 and 330 K.

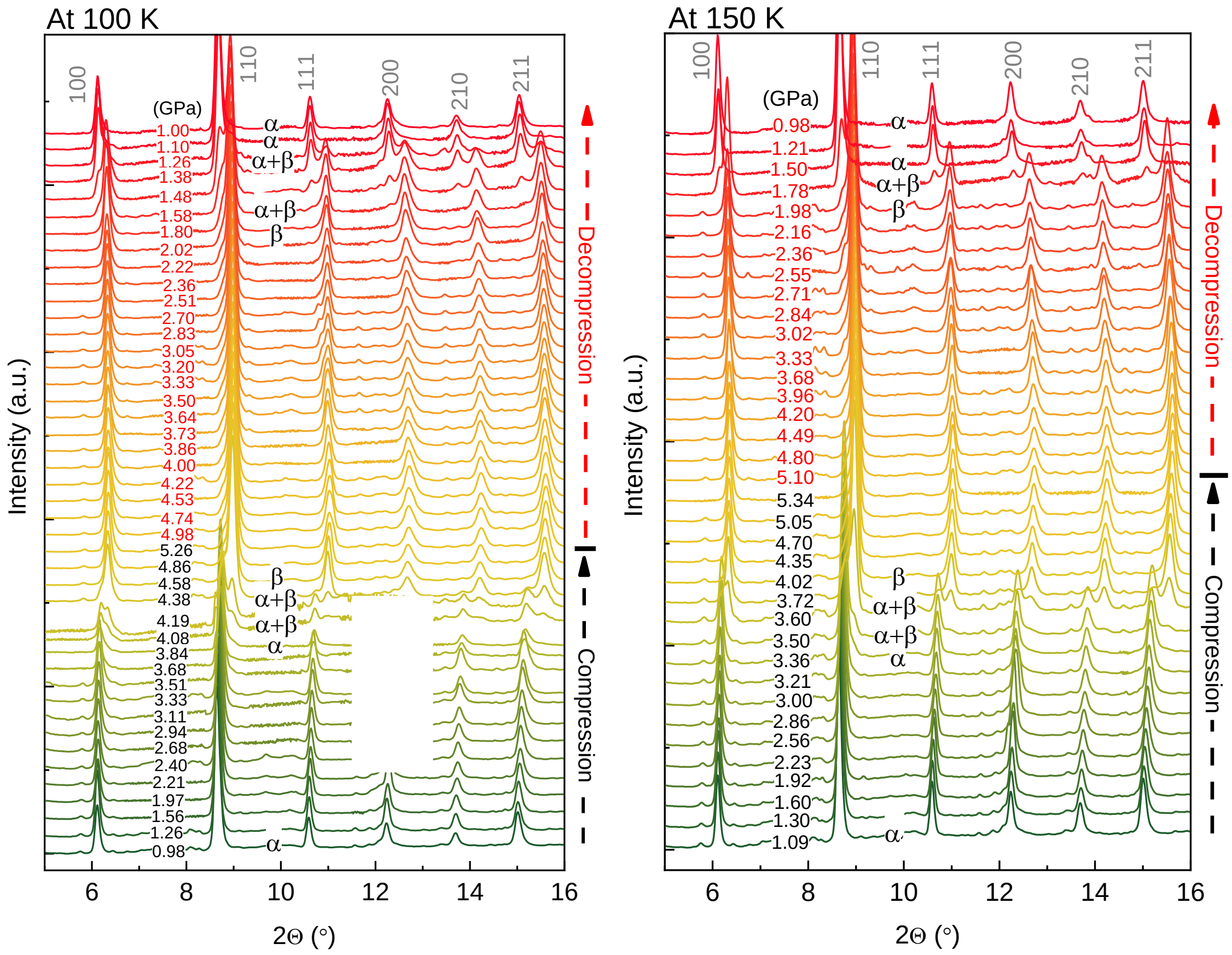

**Fig. S2.** High-P XRD patterns collected at 100 K (left panel) and 150 K (right panel). The masked region in the left panel corresponds to impurity peaks that may arise from the gasket system.

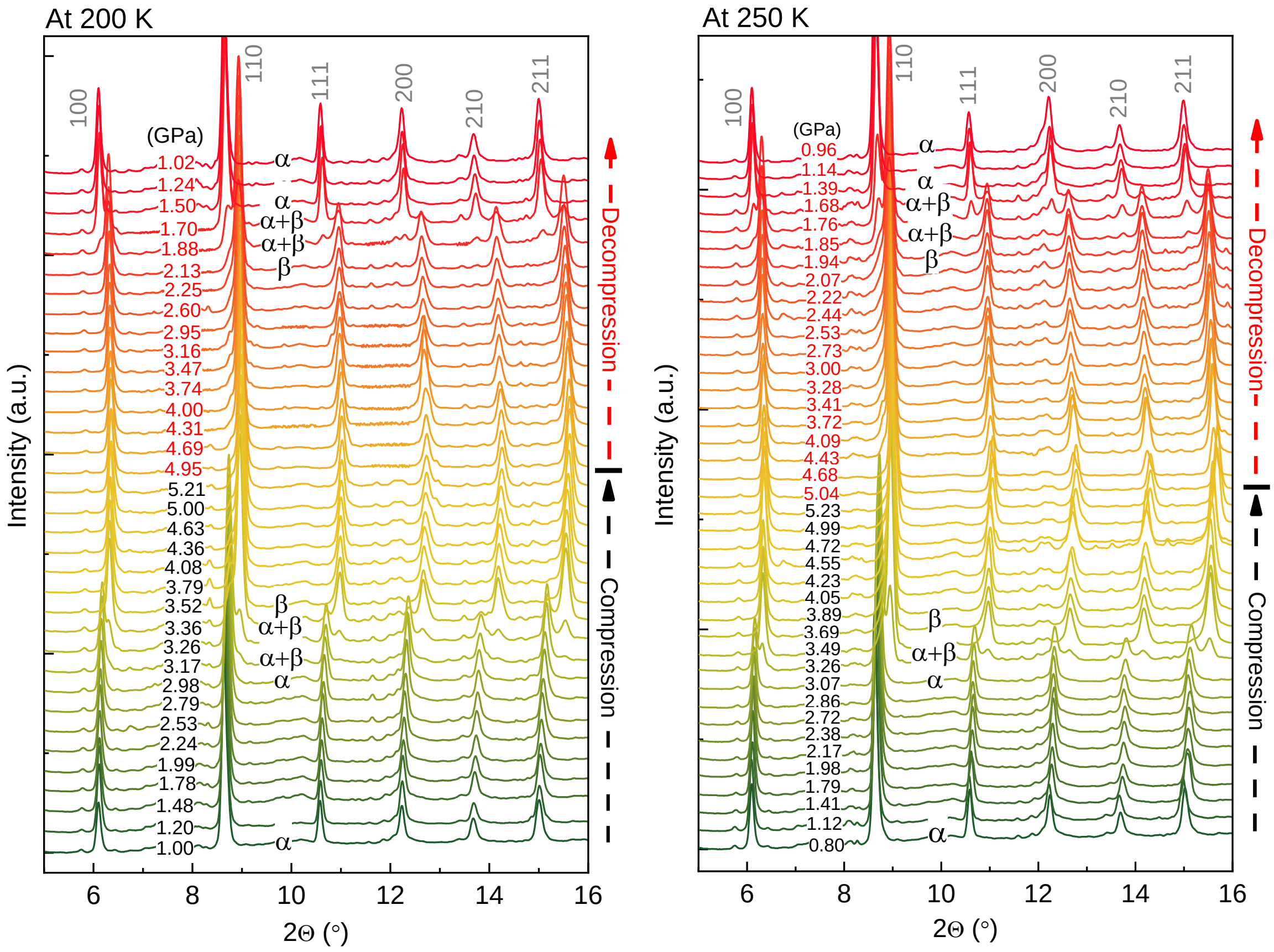

**Fig. S3.** High-P XRD patterns collected at 200 K (left panel) and 250 K (right panel).

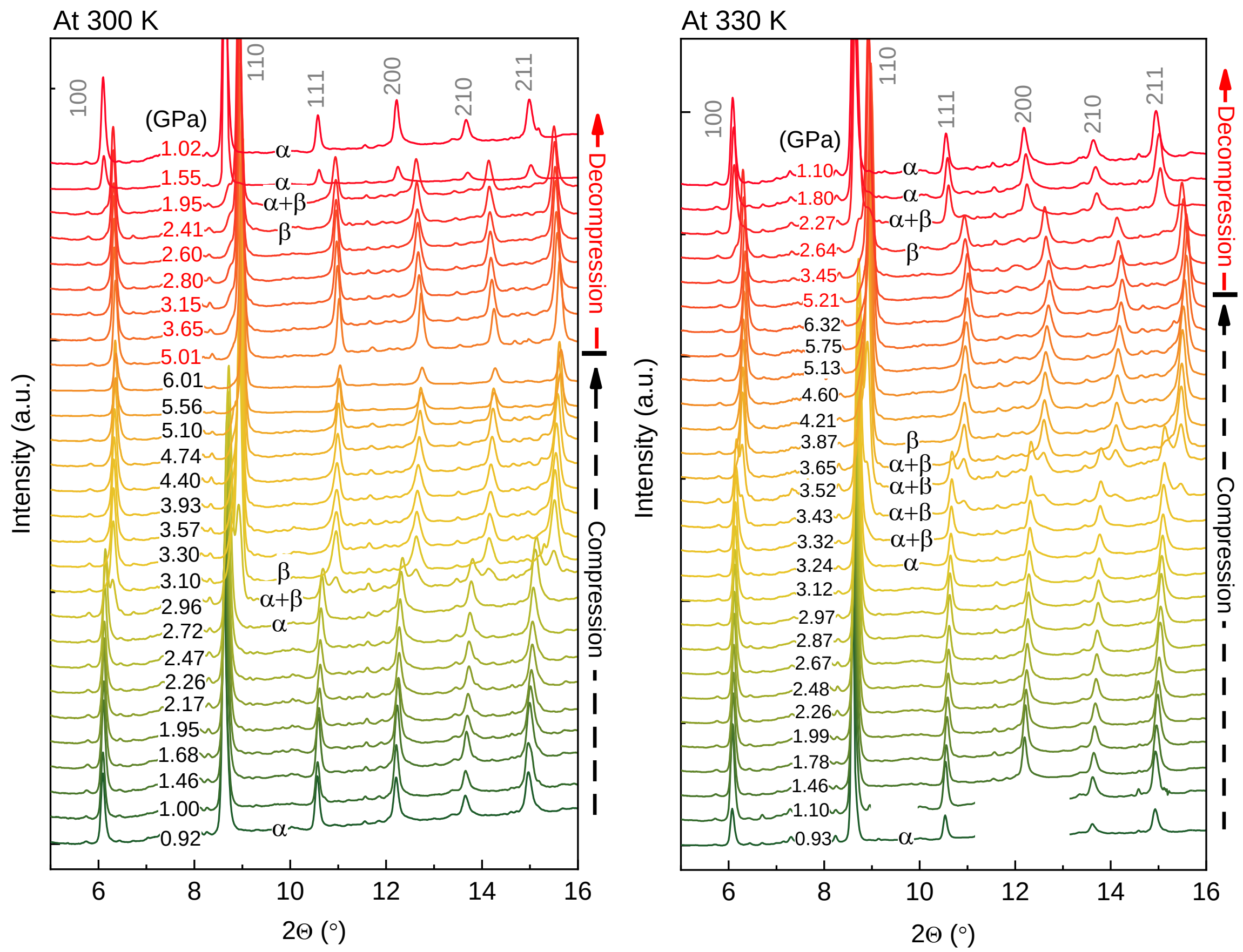

**Fig. S4.** High-P XRD patterns collected at 300 K (left panel) and 330 K (right panel). The masked portion of patterns at 330 K corresponds to the peaks of the gasket system.

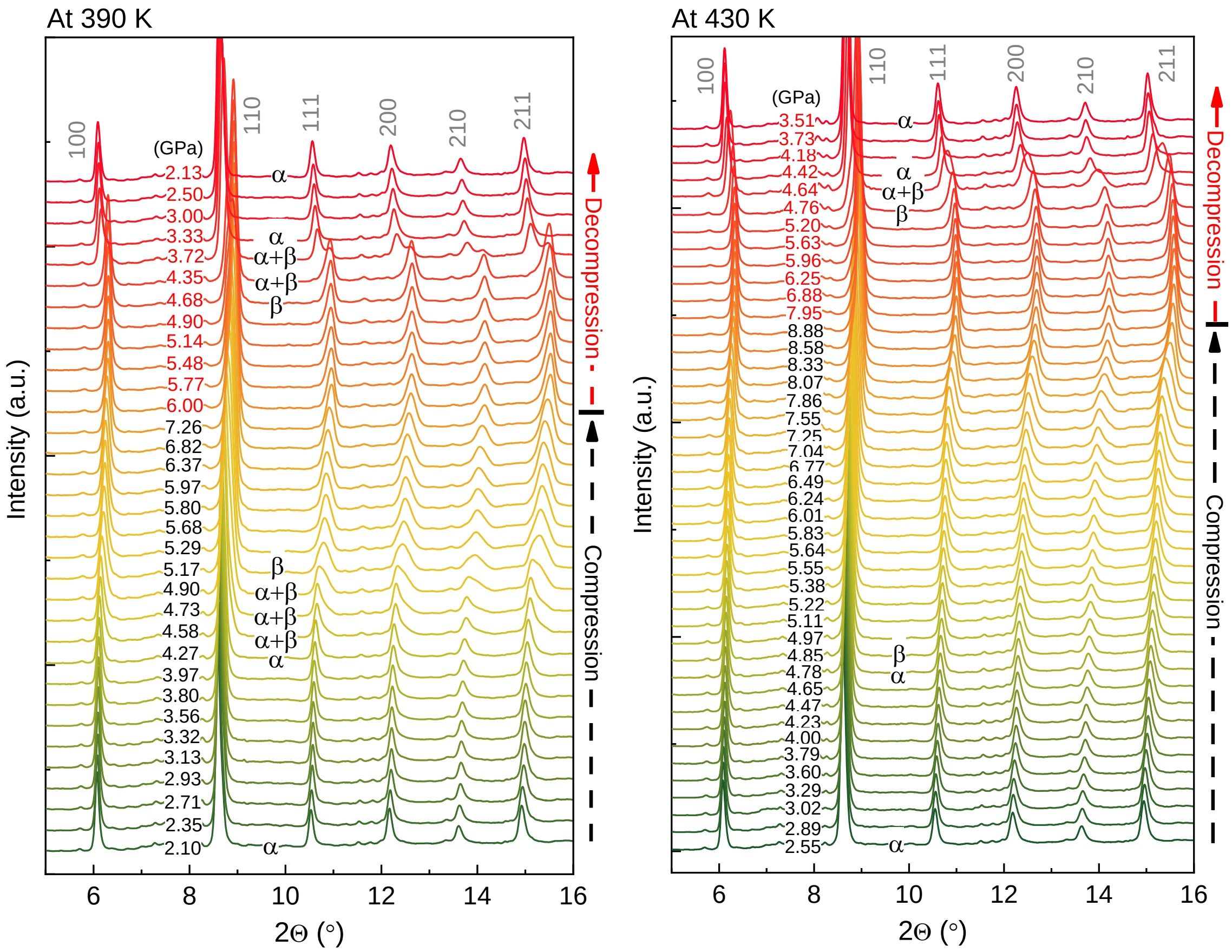

**Fig. S5.** High-P XRD patterns collected at 390 K (left panel) and 430 K (right panel).

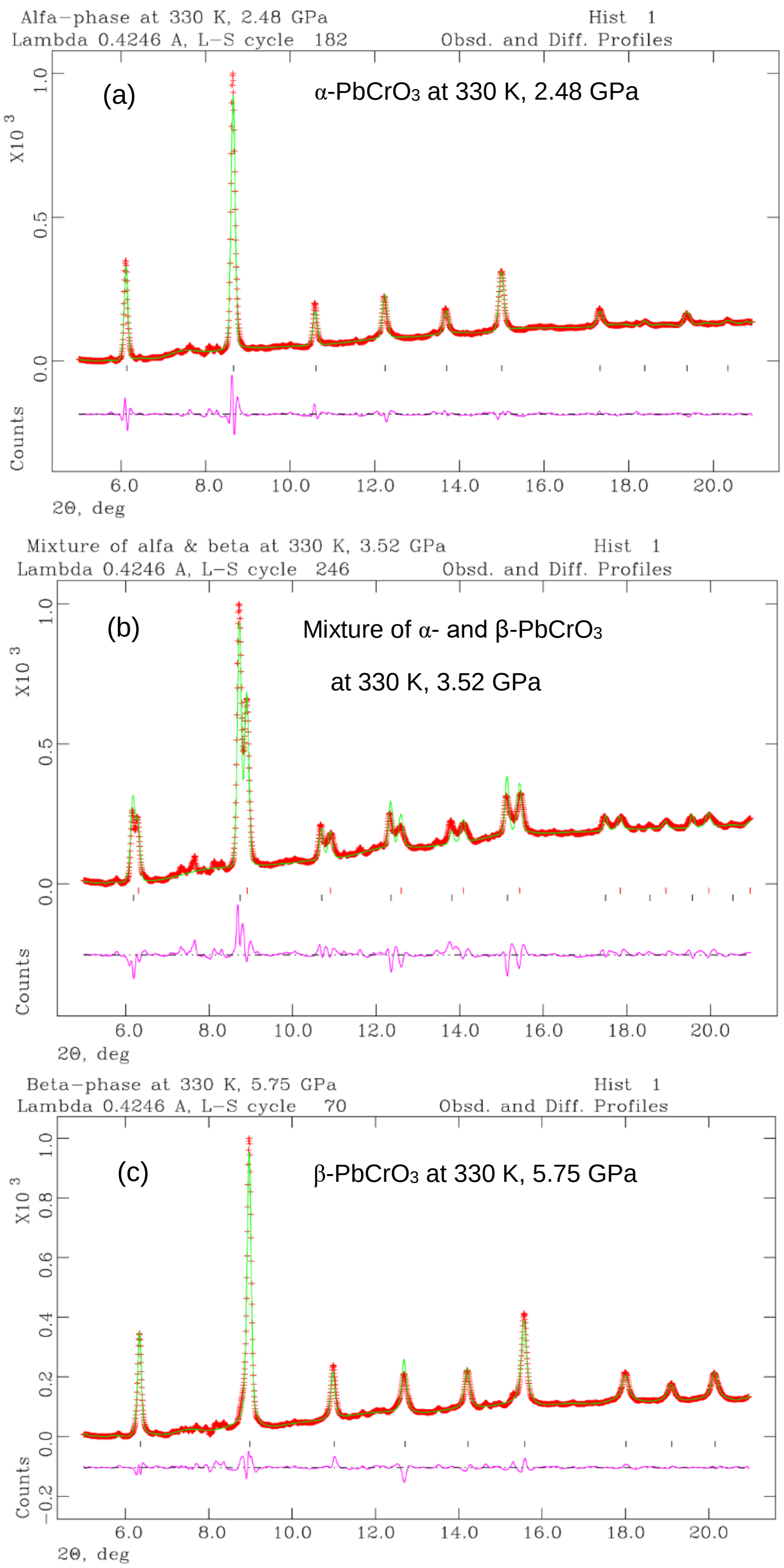

**Fig. S6.** Refined XRD patterns for $PbCrO_3$ taken at 330 K and around the transition pressure. The wavelength of incident synchrotron radiation is 0.4246 Å. (**a**) α-$PbCrO_3$ at 330 K and 2.48 GPa. (**b**) Mixture of α- and β-$PbCrO_3$ at 330 K and 3.52 GPa. The refined phase volume ratio is α:β = 1.44:1, close to the estimated value in terms of their peak intensity (e.g., 1.55:1 for their 110 lines), which provides a quick check of phase percentage.

(**c**) β-$PbCrO_3$ at 330 K and 5.75 GPa. In each panel, the red crosses and green line represent the observed and calculated, respectively, with the difference between them as denoted by the bottom magenta line. The vertical sticks correspond to the calculated peak positions. The impurity peaks around 2θ = 8º arise from a byproduct of $Pb_5CrO_8$ produced during the high-P synthesis[1].

The refinements were performed using the GSAS program. Before the data refinement, the GSAS readable format of data were converted from the regular 2θ-intensity data using the CMPR program. The detailed procedures of refinements can be found elsewhere[1].

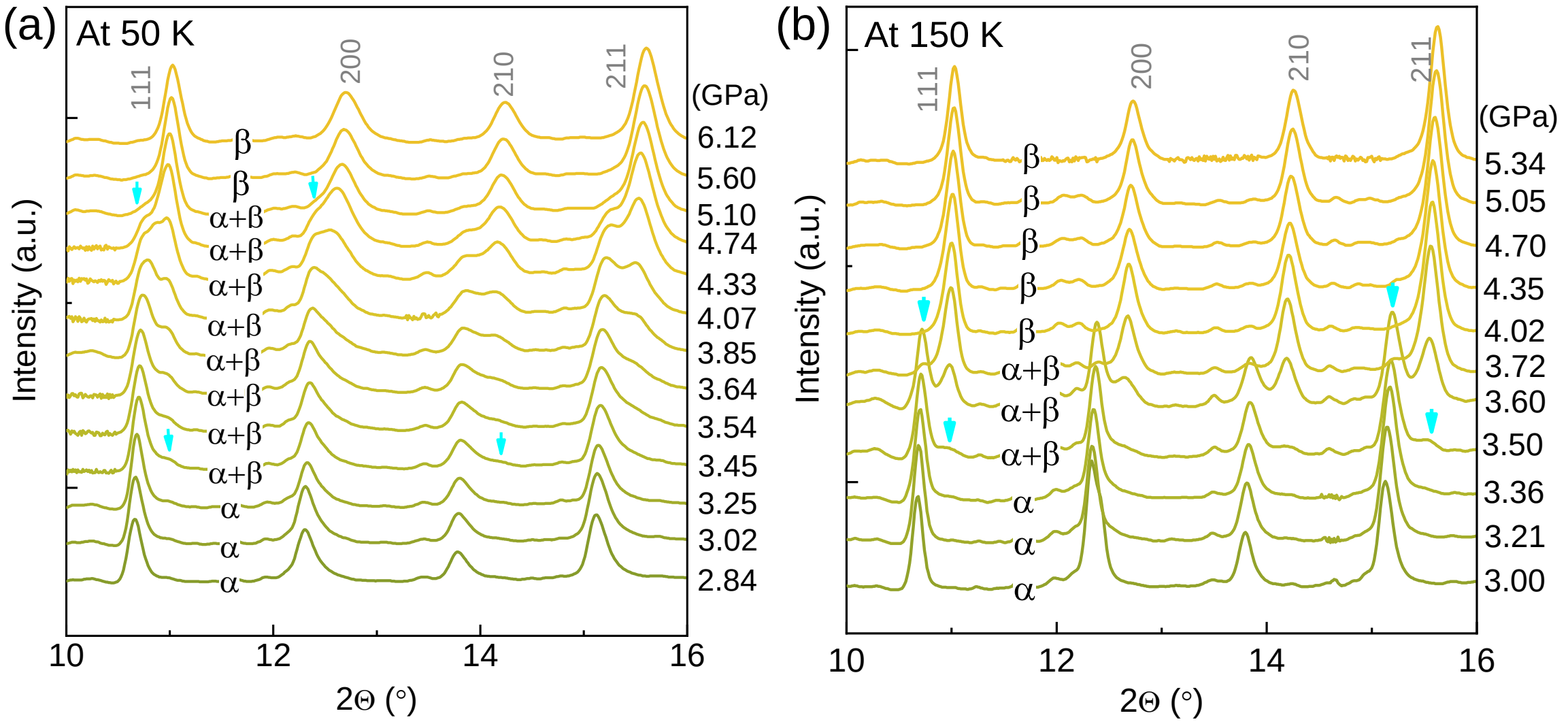

**Fig. S7.** Enlarged portions of XRD patterns taken on isothermal compression at 50 and 150 K, respectively, to show the details of phase evolution with varying pressure. The cyan arrows denote where the new phase comes out and disappears, based on which the onset and ending of phase transition can be discerned. (**a**) At 50 K. The onset α→β transition occurs at 3.45 GPa and completes at 5.10 GPa. (**b**) At 150 K. The onset α→β transition occurs at 3.50 GPa and completes at 4.02 GPa.

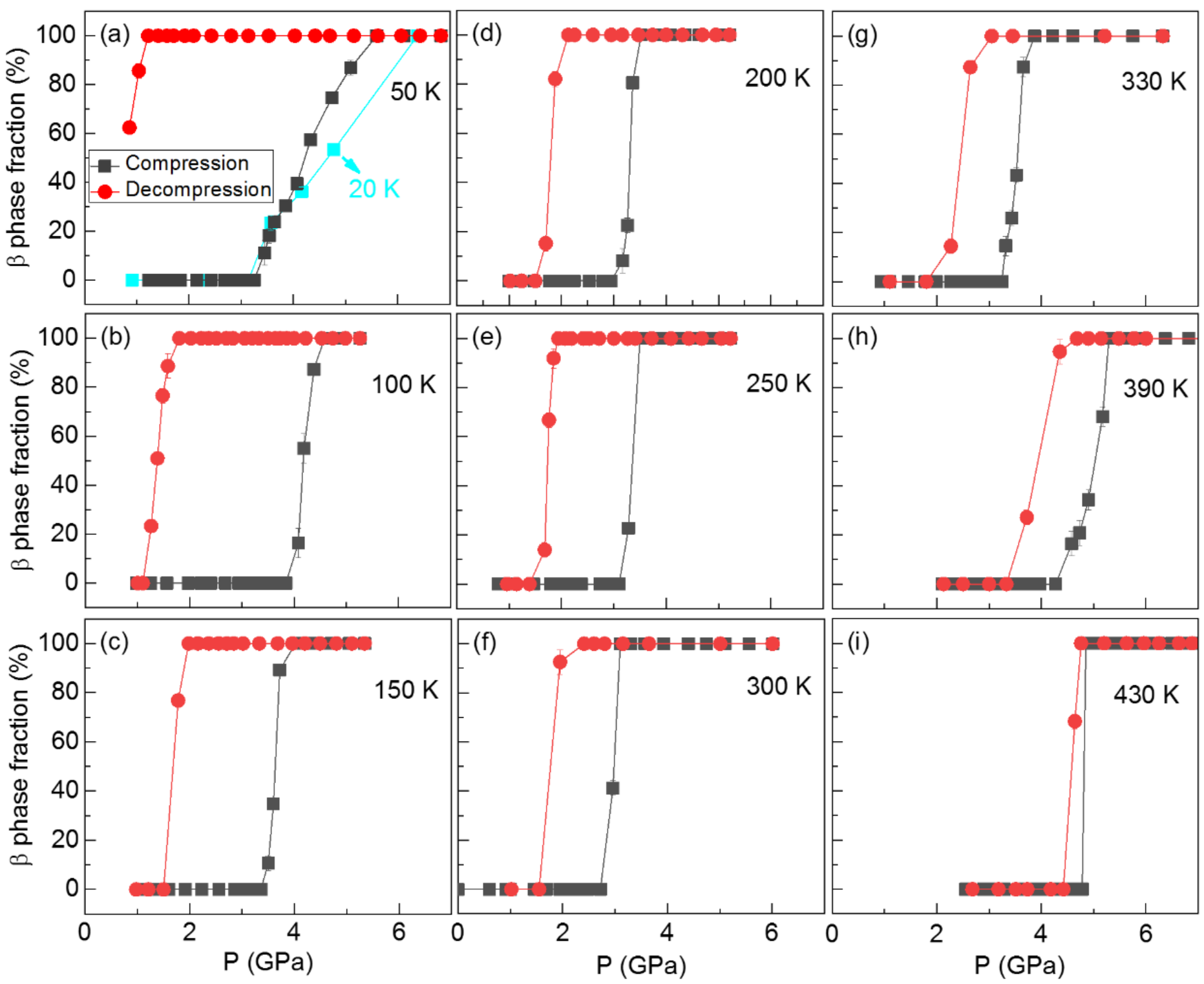

**Fig. S8.** Refined phase fraction for β-$PbCrO_3$ during the isothermal compression (black squares) and decompression (red dots), based on the analysis of isothermal high-P XRD data. (**a**) 50 K. The case of compression at 20 K is also plotted in cyan. (**b**) 100 K. (**c**) 150 K. (**d**) 200 K. (**e**) 250 K. (**f**) 300 K. (**g**) 330 K. (**h**) 390 K. (**i**) 430 K.

As expected, the variation of phase fraction against pressure gives rise to isothermal hysteresis loops. At 50 K, although the reverse transition is incomplete, the hysteresis loop is markedly large, because the transition is kinetically more sluggish at low temperatures. As the temperature increases, the hysteresis loop gradually decreases and almost diminishes at $T_c$, which signals a crossover from the first to second order transition, similar to that of the observed from P-V data (see Fig. 1 of main text).

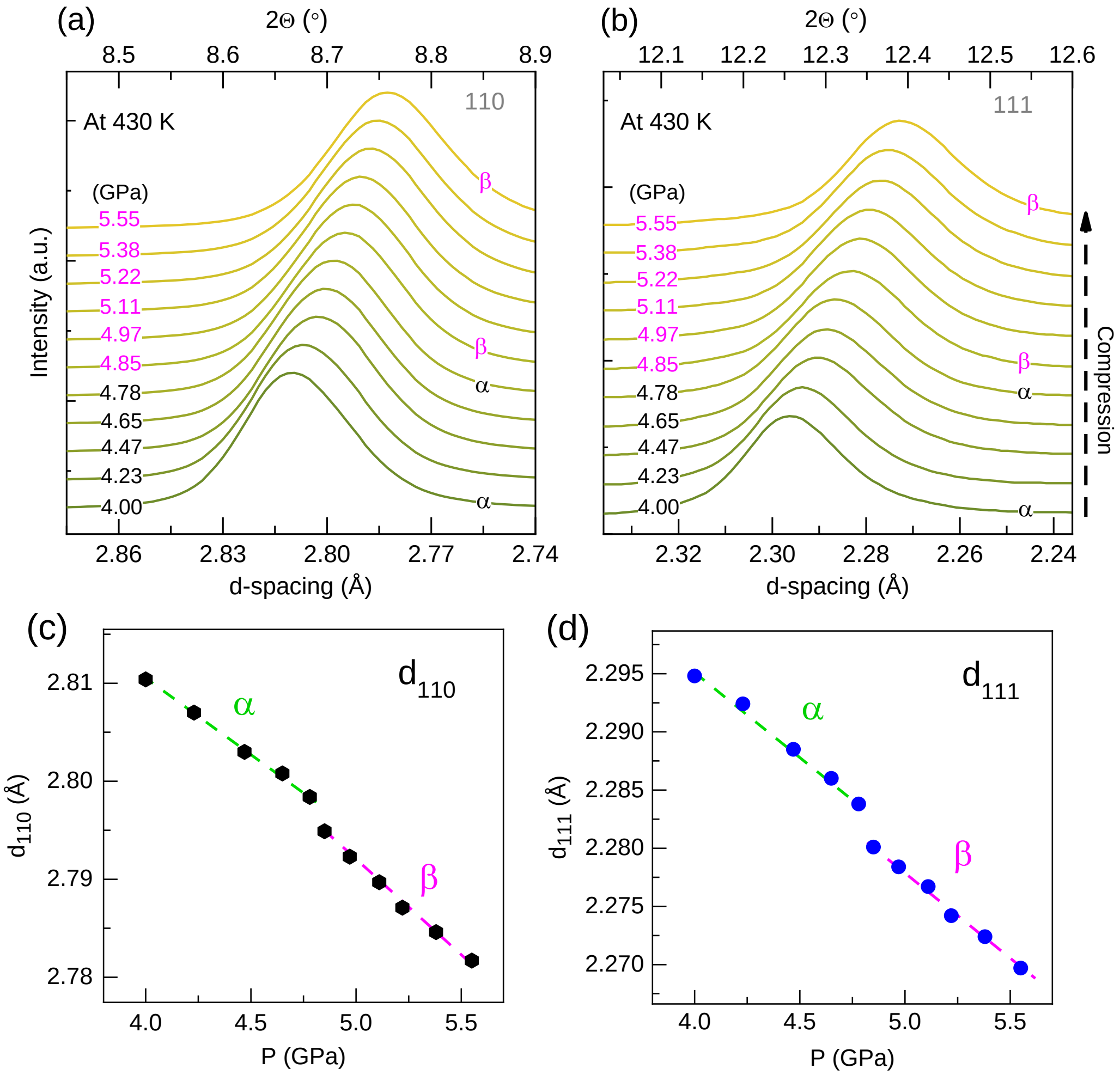

**Fig. S9.** Evolution of peak profile and position of the 110 and 111 lines at 430 K during the isothermal compression around $P_c$ = 4.82 GPa. (**a**) - (**b**) Enlarged portions of the 100 and 200 diffraction lines. Their full patterns are given in Fig. S5. (**c**) - (**d**) Derived d-spacings as a function of pressure. The error bars are too small to show.

With increasing pressure, both the 110 and 111 lines gradually shift to the low d-spacing side without involving the occurrence of new peaks or peak split. Besides, pressure also induces a slightly-increased peak broadening. On a close look of the d-spacing values in (c) - (d), a clear $\alpha \rightarrow \beta$ transition can be identified around $T_c$. Other peaks also have a similar d-spacing variation around $T_c$, which agrees well with that of the observed from our P-V data (see Fig. 1(k) of the main text).

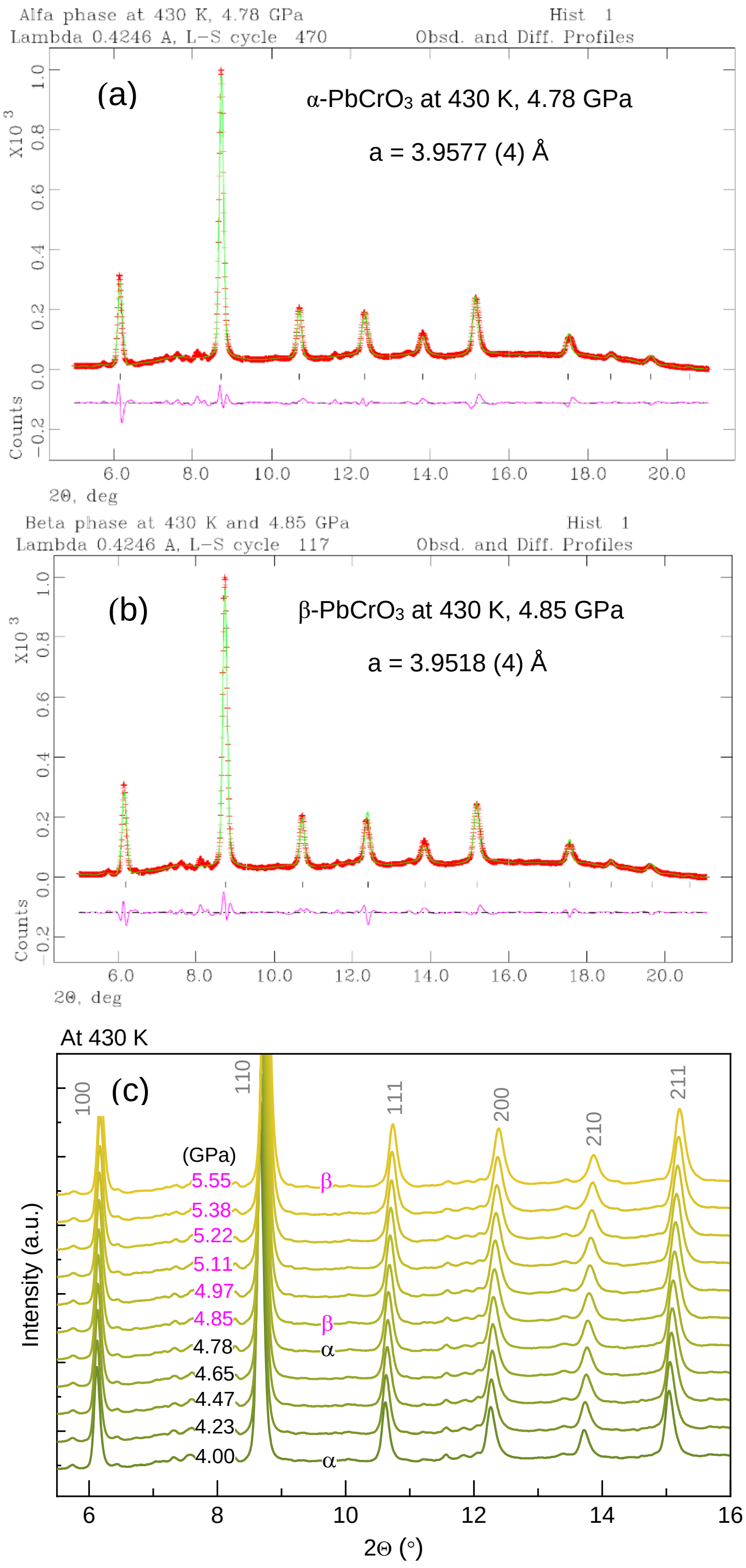

**Fig. S10.** (**a**) - (**b**) Refined XRD patterns for α- and β-$PbCrO_3$ taken at 430 K and two different pressures of 4.78 and 4.85 GPa, respectively. (**c**) Selected XRD pattern taken at 430 K and pressures of 4 - 5.55 GPa to show more details.

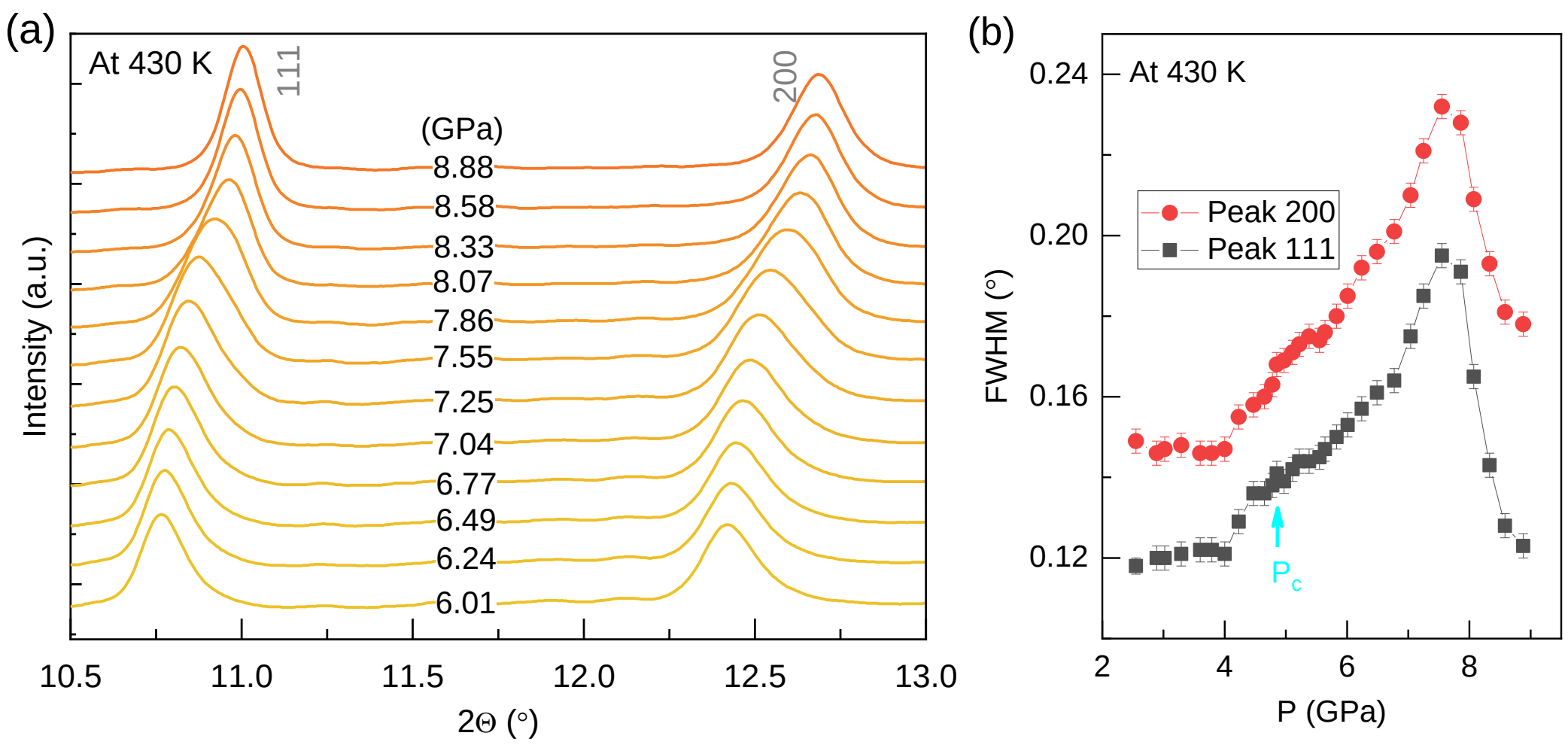

**Fig. S11.** (**a**) Evolution of peak broadening of the 111 and 200 lines at 430 K and in the 6.01 - 8.88 GPa pressure range. (**b**) Derived the full width at half maximum (FWHM) of selected peaks of 111 and 200 as a function of pressure, based on analysis of XRD data using the JADE program.

Anomalous peak broadening can be readily identified in Fig. S11 (a), especially at 7.55 GPa. To quantitatively demonstrate width of the broadened peaks, we plot their FWHM values in Fig. S11(b). Apparently, with the increase of pressure, the widths of both peaks have a similar trend of increase and reach their maximum values of ~0.232 and 0.195$^{o}$ around 7.6 GPa, respectively, more than 65% and 55% larger than those of their low-P widths with values of ~0.118 and 0.149$^{o}$. On a note, each peak in Fig. S11(b) also shows a clear broadening before the $\alpha \rightarrow \beta$ transition at $P_c$.

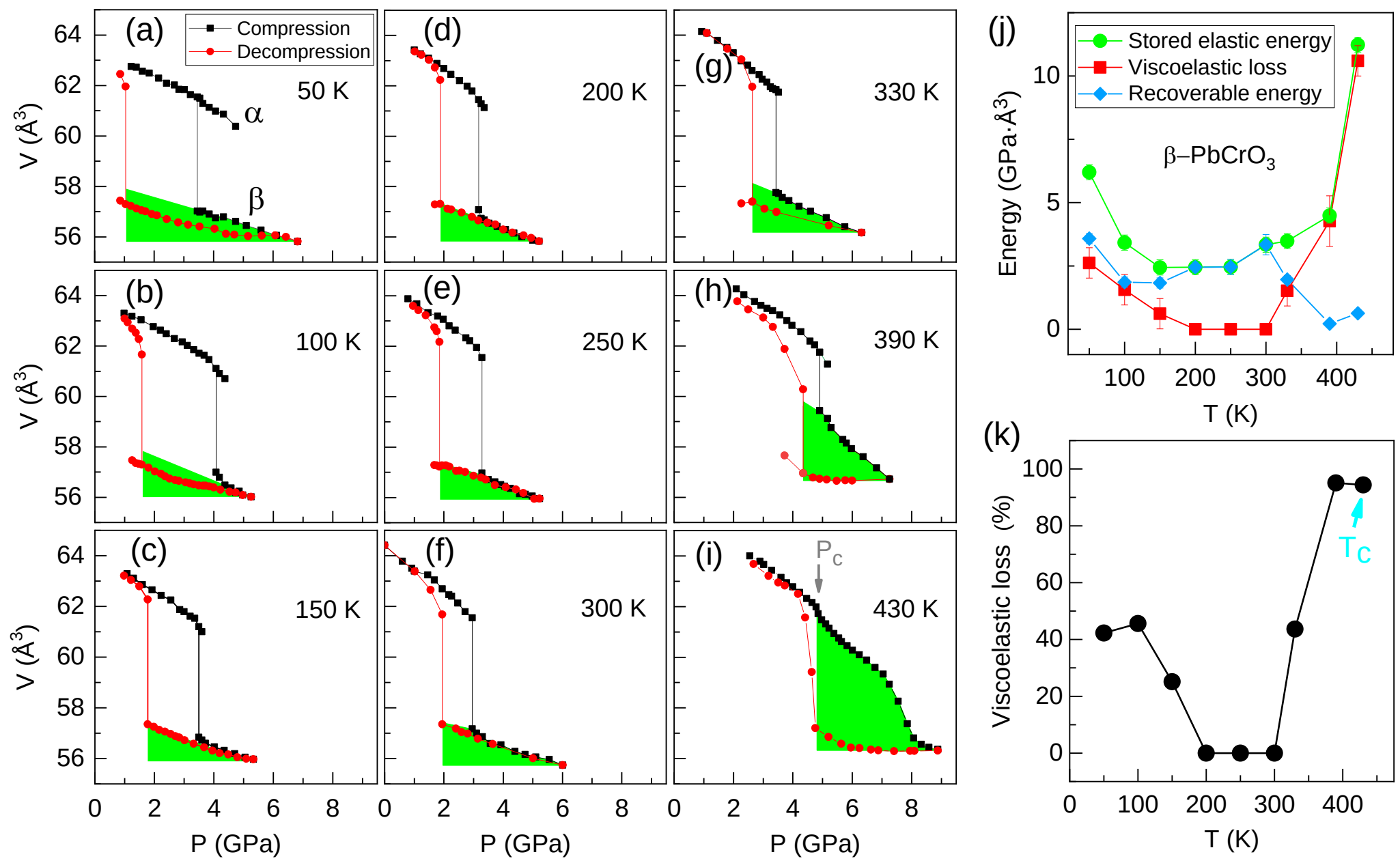

**Fig. S12.** (**a**)-(**i**) Isothermal P-V data with integrated areas for the calculations of total elastic energy storage in β-$PbCrO_3$ at the end of each compression, as denoted in green. (**j**) Derived total elastic energy *vs.* temperature (green line). Also plotted is the variations of viscoelastic loss and recoverable elastic energy against temperature, respectively, as denoted in red and blue. See Fig. 1 of the main text for the calculation of viscoelastic loss. The recoverable elastic energy can also be calculated from the difference between the total elastic energy and viscoelastic loss. (**k**) Ratio of viscoelastic loss to total elastic energy storage in the β phase at the end of each compression.

The viscoelastic effect can be quantitatively defined by the ratio of viscoelastic loss to total elastic energy storage in β-$PbCrO_3$ at the end of each compression, as shown in Fig. S12. The total elastic energy stored in β-$PbCrO_3$ can be assessed by the integration of each compression P-V area (green areas in Figs. S12(a) - S12(i)). The thus-obtained elastic energy, viscoelastic loss, and recoverable energy are plotted in Fig. S12(j) as a function of temperature. Fig. S12(k) shows the percentage of viscoelastic loss during decompression of the β phase (i.e., the ratio of viscoelastic loss to elastic energy). Obviously, the phase remains nearly elastic in the 200 - 300 K range, while the viscoelastic loss increases with either the decrease or increase of temperature, indicating enhancements of viscoelasticity. In particular, the stored elastic energy is nearly completely dissipated due to the viscoelastic effect in the vicinity of $T_c$, corresponding to a *giant or super* effect of viscoelasticity. It is noted that more than 40% elastic energy is dissipated during the decompression of the β phase below 100 K, which further indicates the 'liquid' state of electrons with a large viscosity.

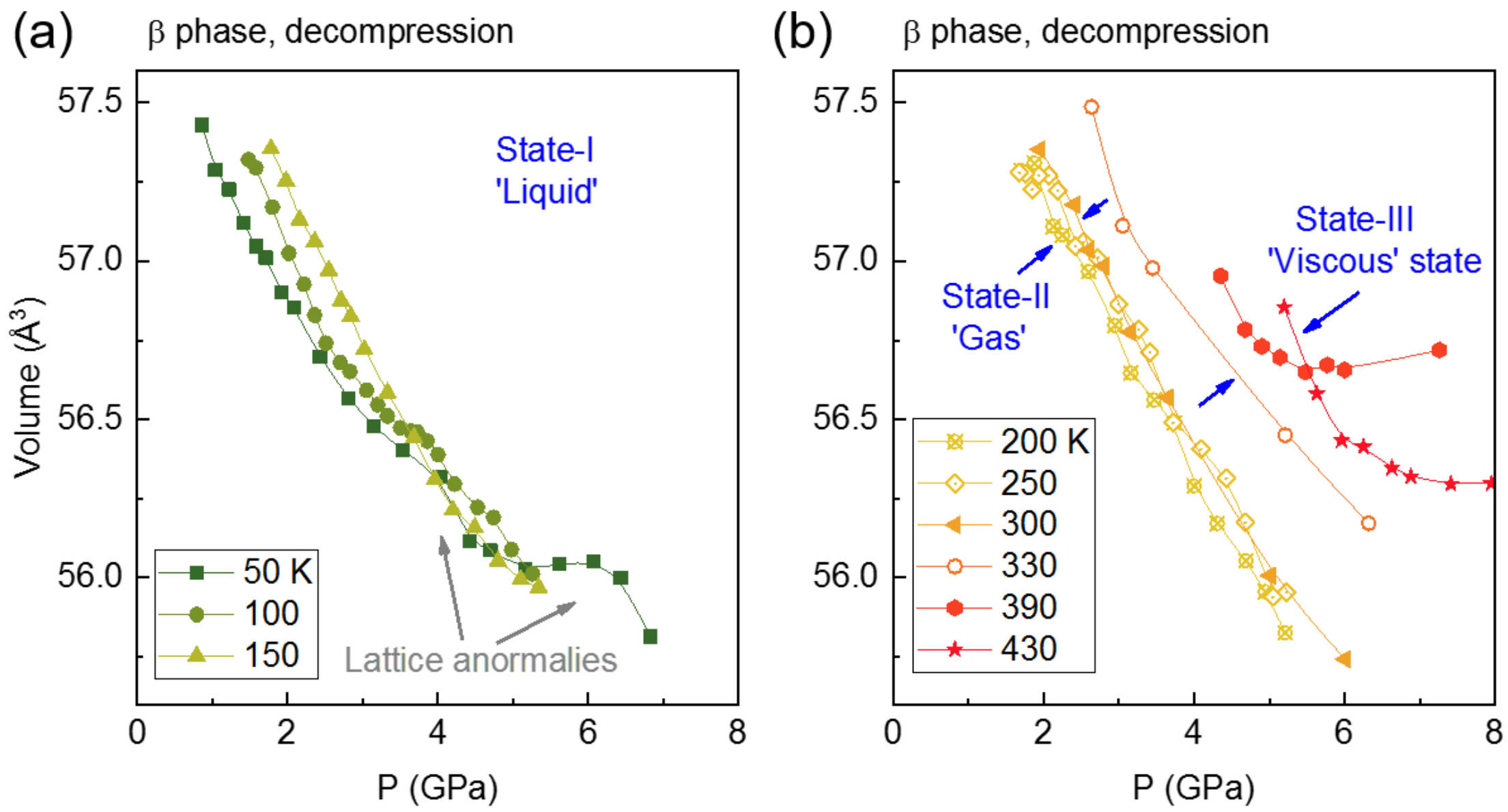

**Fig. S13.** Responses of the lattice to different electronic states in the metallic β-$PbCrO_3$ during decompression at various temperatures. (**a**) State-I at 50 - 150 K. The delocalized electrons in β-$PbCrO_3$ at 50 - 150 K may behave like "liquid" with certain viscosity above 2 GPa, leading to obvious lattice anomalies around 2 - 6 GPa and 50 K. Such lattice anomalies gradually disappear above 150 K. (**b**) State-II at 200 - 300 K and State-III above 300 K. Clearly, the isothermal P-V data at 200 - 300 K are nearly linear, which can be well described by 2$^{nd}$ Birch-Murnaghan equations of state (EoS). This indicates that the delocalized electrons in this temperature regime are gas-like and weakly interact with the lattice degree of freedom. Around $T_c$ = 430 K, giant lattice anomalies are observed as a result of large viscoelastic effects.

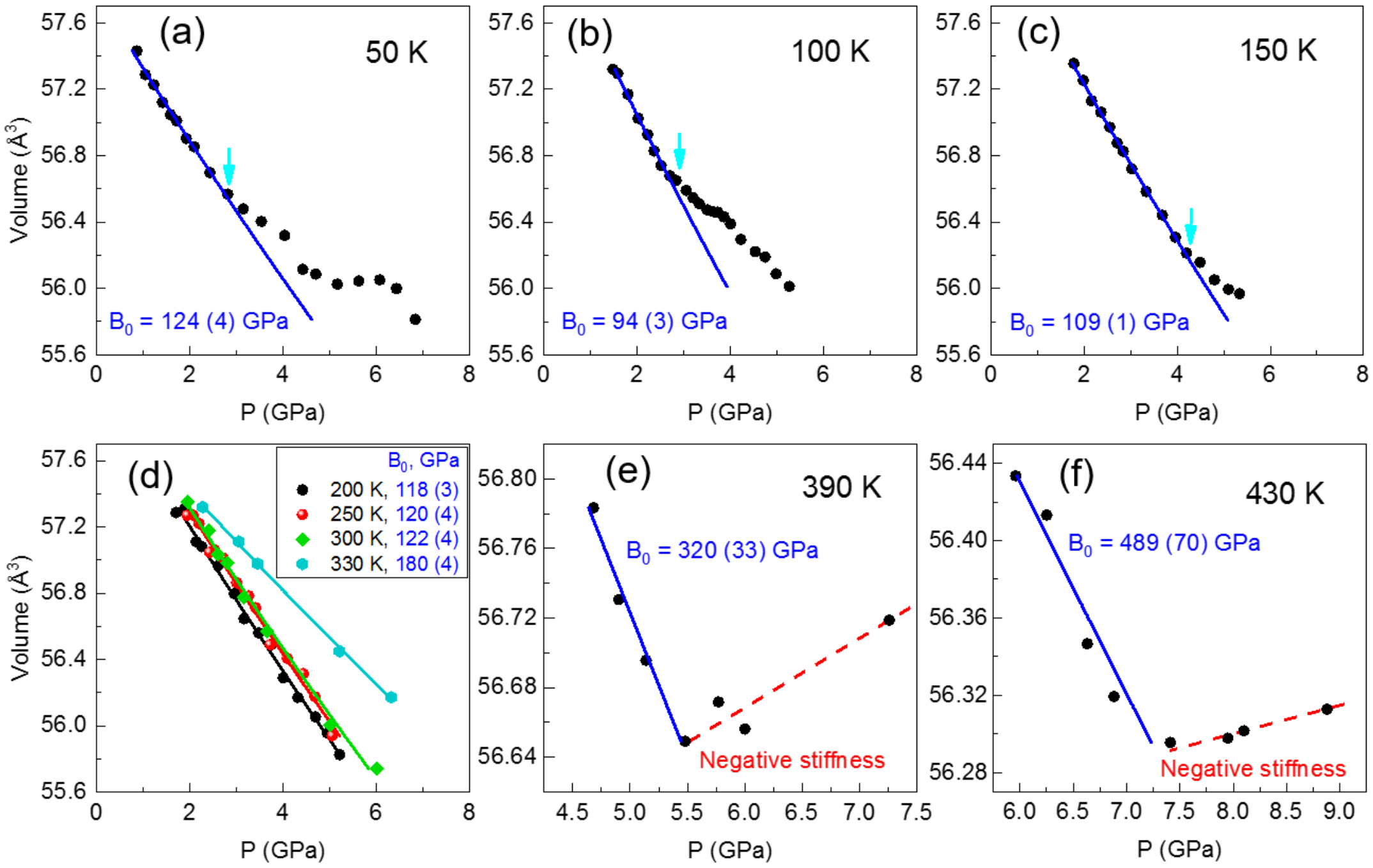

**Fig. S14.** P-V data for β-$PbCrO_3$ collected during decompression at different temperatures. (**a**) 50 K. (**b**) 100 K. (**c**) 150 K. (**d**) 200 K, 250 K, 300 K, and 330 K. (**e**) 390 K. (**f**) 430 K. Solid dots represent the observed data for β-$PbCrO_3$. At each temperature, the fit of the associated P-V data to a $2^{nd}$ Birch-Murnaghan EoS gives rise to the associated bulk modulus ($B_0$), as denoted by solid lines.

In (a)-(c), at higher pressures the volumes have a clear deviation from their EoS lines, as a result of the viscoelastic effect that disappears at 200 - 300 K, leading to nearly identical $B_0$ value (i.e., ≈120 GPa). Above 300 K, the obtained $B_0$ promptly increases upon approaching to 430 K. Remarkably, the negative stiffnesses occur at 390 K and 430 K at the early stage of decompression, followed by an extremely large dynamic bulk modulus of ~320 GPa and 489 GPa, respectively. At $T_c$ = 430 K, the determined dynamic $B_0$ (i.e., 489 GPa) is even much greater than that of diamond (i.e., 440 GPa), indicating super-large viscoelasticity near the critical endpoint.

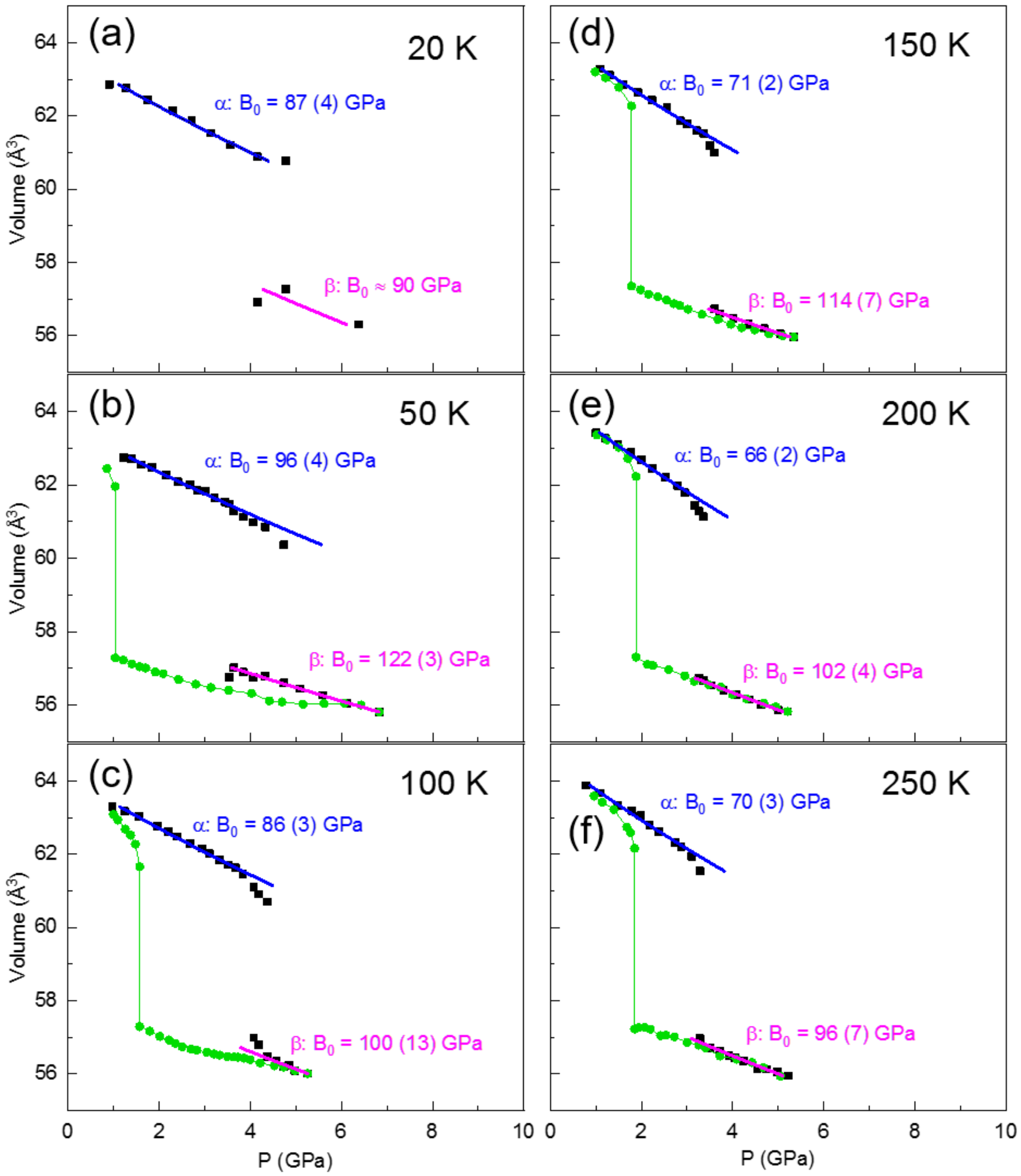

(a)
20 K
α: $B_0$ = 87 (4) GPa
β: $B_0$ ≈ 90 GPa
(b)
50 K
α: $B_0$ = 96 (4) GPa
β: $B_0$ = 122 (3) GPa
(c)
100 K
α: $B_0$ = 86 (3) GPa
β: $B_0$ = 100 (13) GPa
(d)
150 K
α: $B_0$ = 71 (2) GPa
β: $B_0$ = 114 (7) GPa
(e)
200 K
α: $B_0$ = 66 (2) GPa
β: $B_0$ = 102 (4) GPa
(f)
250 K
α: $B_0$ = 70 (3) GPa
β: $B_0$ = 96 (7) GPa
Volume (Å$^3$)
P (GPa)

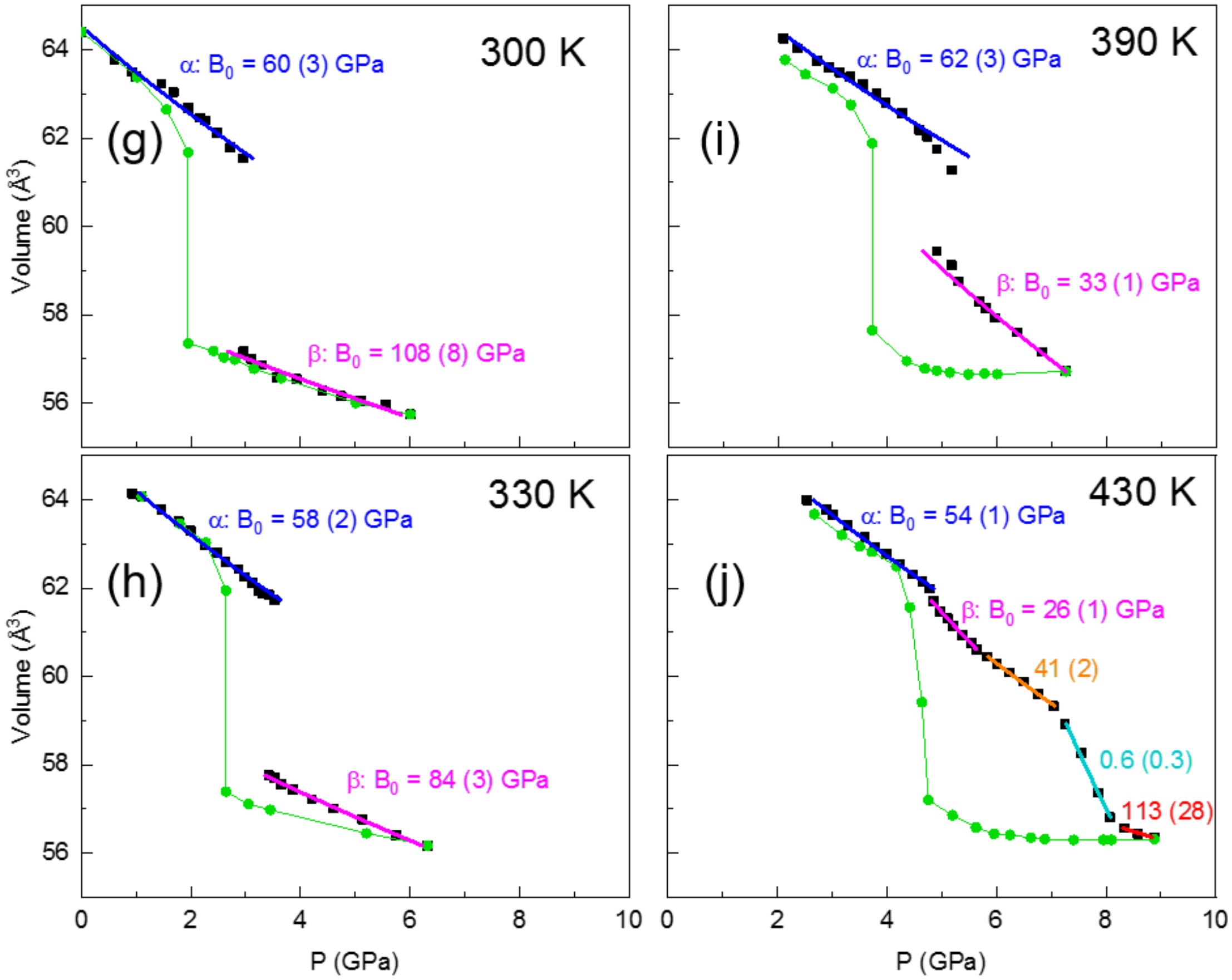

**Fig. S15.** Calculations of bulk moduli for both α- and β-$PbCrO_3$ using the P-V data collected during isothermal compression at different temperatures of 20 K, 50 K, 100 K, 150 K, 200 K, 250 K, 300 K, 330 K, 390 K, and 430 K. In each panel, the blue and magenta lines represent the fits of P-V data of the α and β phases (black squares) to 2$^{nd}$ order Birch-Murnaghan EoS, respectively. The green dots denote the P-V data taken during decompression, which are plotted for comparison.

Note that the data taken on the thresholds of phase transition deviate from the EoS line, which can be attributed to the lattice strain as a result of volume reduction or expansion during transition[6]. To accurately determine the bulk modulus, such deviated data points are excluded for achieving appropriate analysis. It is noted that the determined bulk modulus for β-$PbCrO_3$ at 300 K is $B_0$ = 108 (8) GPa, which is smaller than that of previous report with a value of ~187 GPa (fixed $B_0$’ = 4)[7]. The data of the latter was obtained based on both the compression and decompression in a wide pressure range of 3 - 30 GPa, so that the viscoelastic effect should be involved, leading to an enhanced value. In fact, using the 3 - 20 GPa data, we have also determined an intermediate value of $B_0$ = 148 (4) GPa (fixed $B_0$’ = 4), confirming our speculation.

At $T_c$ = 430 K, the P-V curve of β-$PbCrO_3$ can be divided into four pressure regions of 4.8 - 5.8 GPa, 5.8 - 7.2 GPa, 7.2 - 8.2 GPa, and 8.2 - 9 GPa. The thus-estimated bulk moduli for these regions are 26 (1), 41 (2), 0.6 (0.3), and 113 (28) GPa, respectively.

**Table S1.** Summaries of bulk moduli of both α- and β-$PbCrO_3$ calculated from the equation of state (see Figs. S14 - S15). The pressure ranges that are used for calculations of bulk modulus are also included in the brackets aside.

| | **Bulk modulus/GPa, α-$PbCrO_3$** | **Bulk modulus/GPa, β-$PbCrO_3$** | |
|---|---|---|---|
| T/K | Compression | Compression | Decompression |
| 20 | 87(4) (0.9-4.2 GPa) | ≈90 (4.1-6.4 GPa) | - |
| 50 | 96(4) (1.3-3.8 GPa) | 122(3) (3.6-6.9 GPa) | 124(4) (0.8-2.8 GPa) |
| 100 | 86(3) (1.0-3.9 GPa) | 100(13) (4.4-5.3 GPa) | 94(3) (1.4-2.8 GPa) |
| 150 | 71(2) (1.1-3.4 GPa) | 114(7) (3.6-5.3 GPa) | 109(1) (2.3-5.3 GPa) |
| 200 | 66(2) (1.0-3.0 GPa) | 102(4) (3.3-5.3 GPa) | 118(3) (1.7-5.2 GPa) |
| 250 | 70(3) (0.8-3.1 GPa) | 96(7) (3.2-5.2 GPa) | 120(4) (1.9-5.4 GPa) |
| 300 | 60(3) (0.1-3.0 GPa) | 108(3) (3.0-5.6 GPa) | 122(4) (1.8-5.9 GPa) |
| 330 | 58(2) (0.9-3.5 GPa) | 84(3) (3.5-6.3 GPa) | 180(4) (2.2-6.3 GPa) |
| 390 | 62(3) (2.1-4.8 GPa) | 33(1) (5.3-7.3 GPa) | 320(33) (1.3-3.8 GPa) |
| 430 | 54(1) (2.5-4.8 GPa) | 26(1) (4.9-5.7 GPa)<br>42(2) (5.8-7.0 GPa)<br>0.6(0.3) (7.3-8.1 GPa)<br>113(28) (8.3-8.9 GPa) | 489(77) (6.0-7.4 GPa) |

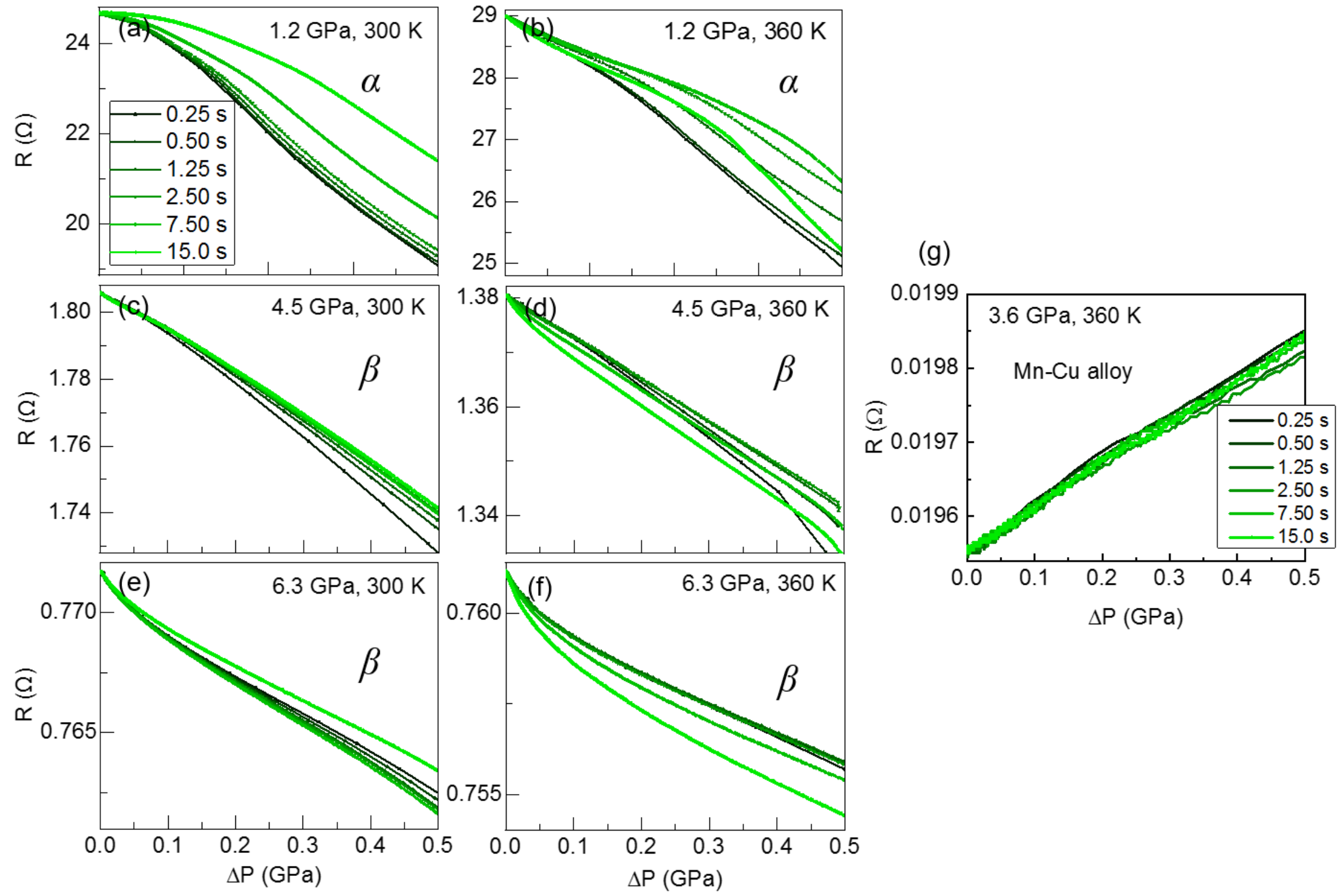

**Fig. S16.** Time-dependent variation of electrical resistance of $PbCrO_3$ with a same pressure increase of ΔP = 0.5 GPa at selected P - T conditions, using a piezo-driven DAC[5]. At each pressure, the data were taken at two different temperatures of 300 and 360 K, respectively, for comparison (also see Experimental details). (**a**) - (**b**) α-$PbCrO_3$ at 1.2 GPa. (**c**) - (**f**) β-$PbCrO_3$ at 4.5 and 6.3 GPa, respectively. (**g**) Measurements on standard material of Mn-Cu alloy to show the experimental uncertainty. The associated data was also used for the pressure calibration. At each desired P-T condition, the pressure was continuously varied in a small pressure range of ΔP = 0 - 0.5 GPa with different loading durations of 0.25, 0.50, 1.25, 2.50, 7.50, 15.0 s, respectively, to check the time dependency of transport behaviors. The electrical resistance data were collected during multiple compression-decompression cycles, but only one cycle of compression data was picked up for the study at each target P-T condition. Because of the limited working temperature for piezoelectric element, the experimental temperature was restricted below 360 K to prevent the breakdown of piezoelectric ceramics.

In principle, the lattice of a material can be directly changed by external pressure, while the response of charges (or electrons) to such lattice deformation often leads to the variation of electrical transport properties. In normal materials without involving electron viscosity, the electrons should have a fast respond to lattice deformation at pressure. However, the appearance of electron viscosity in the material is expected to produce viscoelasticity, which would result in unusual deformations and hence the anomalous time-dependent effects at pressure. Accordingly, the rate- or time-dependent electrical transport behaviors can be well explained.

In Figs. S16(a) - S16(b), $PbCrO_3$ is in an insulating state at 1.2 GPa with thermally-excited charges and there is no viscoelasticity involved, showing a normal time-dependent variation of resistance with pressure. Clearly, a faster increase of pressure corresponds to a faster decreasing of resistance, indicating a prompt response of the lattice to pressure, although the situation at 360 K has slight changes for the 15.0 s line. The normal variation of resistance is also occurred at 4.5 GPa and 300 K (Fig. S16(c)). However, as the P-T conditions approach to the critical endpoint, the time-dependent trend of resistance variation is reversed at 4.5 GPa and 360 K (Fig. S16(d)), and the slower processes (e.g., the 7.50-s and 15.0-s cases) give rise to a faster resistance decrease, likely due to the viscoelastic effect. The similar situation also happens at 6.3 GPa and 300 K. With increasing temperature, this phenomenon is largely enhanced at 360 K, signaling an enhancement of viscoelasticity. It is noted that the above-mentioned time-dependent variation of resistance is intrinsic and not associated with the experimental uncertainty, because our calibration based on the standard Mn-Cu alloy shows little discrepancy with a high accuracy of 0.0001 Ω (Fig. S16(g)).

Apparently, one experimental work cannot solve all the problems when new physics comes up. Future work along this direction is urgently warranted to explore more details of time-dependent effect related to the viscoelasticity of Mott systems.